\documentclass[pra,aps,twocolumn,10pt,superscriptaddress,notitlepage,longbibliography]{revtex4-2}

\usepackage[dvipsnames]{xcolor}
\usepackage[english]{babel} 
\usepackage{graphicx}
\usepackage{float}
\usepackage{braket}
\usepackage{epstopdf}
\usepackage{mathtools}
\usepackage{amsmath}
\usepackage{amssymb}
\usepackage{bbm,bm}
\usepackage[caption=false]{subfig}
\usepackage{pythonhighlight}
\usepackage{hyperref}
\usepackage{float}
\usepackage{bbold}
\usepackage[T1]{fontenc}

\usepackage[markup=underlined]{changes}
\makeatletter
\@namedef{Changes@AuthorColor}{magenta}
\colorlet{Changes@Color}{magenta}
\makeatother
\usepackage{hyperref}
 \hypersetup{
     colorlinks=true,
     linkcolor=blue,
     filecolor=blue,
     citecolor = magenta,      
     urlcolor=red,
     }

\usepackage{braket}

\newcommand{\blue}[1]{{\color{black} #1}}

\usepackage{xargs}
\newcommandx{\greencom}[2][1=]
{\todo[inline, color=green!40,#1]{#2}}
\newcommandx{\bluecom}[2][1=]
{\todo[inline, color=blue!40,#1]{#2}}
\newcommandx{\bluemargin}[2][1=]
{\todo[color=blue!40,#1]{#2}}

\usepackage{letltxmacro}
\LetLtxMacro{\ORIGselectlanguage}{\selectlanguage}
\makeatletter
\DeclareRobustCommand{\selectlanguage}[1]{%
  \@ifundefined{alias@\string#1}
    {\ORIGselectlanguage{#1}}
    {\begingroup\edef\x{\endgroup
       \noexpand\ORIGselectlanguage{\@nameuse{alias@#1}}}\x}%
}
\newcommand{\definelanguagealias}[2]{%
  \@namedef{alias@#1}{#2}%
}

\makeatother

\definelanguagealias{en}{english}
\definelanguagealias{EN}{english}
\definelanguagealias{eng}{english}
\definelanguagealias{de}{ngerman}

\graphicspath{{paper_figs/}}

\begin{document}

\title{Quantum dynamics of few-photon pulsed waveguide-QED with a single artificial atom:
 frequency-dependent 
 scattering theory and time-dependent matrix product states}

\author{Sofia Arranz Regidor}
\email{18sar4@queensu.ca}
\affiliation{Department of Physics,
Engineering Physics and Astronomy, Queen's University, Kingston, Ontario, Canada, K7L 3N6}
\author{Matthew Kozma}
\email{24nkr1@queensu.ca}
\affiliation{Department of Physics,
Engineering Physics and Astronomy, Queen's University, Kingston, Ontario, Canada, K7L 3N6}
\author{Stephen Hughes}
\email{shughes@queensu.ca}
\affiliation{Department of Physics,
Engineering Physics and Astronomy, Queen's University, Kingston, Ontario, Canada, K7L 3N6}

\date{\today}

\begin{abstract} 
We present a quantum dynamical study of pulsed few-photon scattering from a single artificial atom, consisting of a two-level system (TLS) or qubit, in a waveguide QED system, directly comparing and contrasting two different quantum theoretical simulation methods: (i) an input-output scattering approach that uses frequency-dependent scattering matrices, and (ii) a matrix product states (MPS) approach, which uses quantum noise operators in 
discretized time bins and a tensor network technique to directly solve the time-dependent observables for the entire quantum waveguide-qubit system. Beginning with pulsed quantum-field excitation, using one-photon and two-photon Fock state pulses, we first show how to compute time-dependent observables with a scattering matrix approach, in terms of frequency integrals that encode the pulse spectrum, including how to extract population dynamics of the excited quantum emitter, as well as the linear and nonlinear contributions. We present solutions for both symmetric and chiral TLS coupling. We then show how to compute the qubit and field observables in a more direct way using MPS, and obtain the characteristic bird-like shape for the two-photon correlation function at two times, which has been observed in recent experiments. We compare and contrast both of these methods, for one-photon and two-photon excitation pulses, and show excellent agreement. We also present a study of the linear and nonlinear contributions, which can easily be calculated using scattering theory, and which show the important role of pulse duration.
Finally, we demonstrate the clear advantages of MPS by easily going to higher $N$-photon excitations, and show selected example population dynamics of up to eight-photon Fock-state pulses, manifesting in clear nonlinear population oscillations during the pulse interaction, similar to classical Rabi oscillations, but with quantum input fields that have a vanishing electric field expectation value. 

\end{abstract}

\maketitle

\section{Introduction}
\label{sec:intro}

The scattering of propagating photons interacting with a single qubit has been broadly studied in waveguide quantum electrodynamics (waveguide-QED), where the qubit or two-level system (TLS) is coupled to a one-dimensional photonic reservoir (waveguide), and input photons
excite the TLS and create strong light-matter interactions~\cite{Hughes2004,Zheng2010,PhysRevA.83.063828,PhysRevLett.104.023602,PhysRevA.102.023702,PhysRevLett.126.023603,PhysRevA.76.062709,PhysRevLett.98.153003,Witthaut_2010,PhysRevLett.113.263604,Calaj2016,PhysRevLett.116.093601}. Waveguide-QED systems offer a promising route to exploit
scalable few-photon quantum nonlinearities for quantum photonic architectures with a wide range of applications in quantum optics and quantum technologies~\cite{PhysRevLett.115.153901,Lodahl2017,PhysRevResearch.4.023082,Mirhosseini2019}.

Experimentally, significant progress has been made towards a better understanding of these waveguide-qubit processes in various materials systems and at different frequencies~\cite{Trschmann2019,10.1063/1.5117888,PhysRevLett.101.113903,Mirhosseini2019,PhysRevLett.115.163603,PhysRevX.2.011014,PhysRevLett.131.103602,PhysRevA.92.063836,PhysRevX.7.031024,PhysRevLett.120.140404,PhysRevLett.131.033606,PhysRevResearch.2.043213}. For example, superconducting 
qubits interacting with microwave photons are coupled to transmission lines or superconducting waveguides, giving rise to interesting interference effects~\cite{Kannan2020,PRXQuantum.4.030326,PhysRevResearch.6.013279,Wang_2022}, as well as waveguide
resonance fluorescence
\cite{Astafiev2010}
and the generation of non-classical states \cite{PhysRevLett.108.263601}. Semiconductor quantum dots (yielding electron-hole pairs) interact with optical photons in semiconductor waveguides with very high efficiency, and excellent versatility in the nanophotonics design, with an tunable chirality~\cite{PhysRevB.75.205437,PhysRevLett.113.093603,Paesani2019,leFeber2015,PhysRevLett.115.153901,Sllner2015,doi:10.1126/sciadv.aaw0297,PhysRevResearch.4.023082,Liu:22,PhysRevLett.117.240501,PhysRevX.10.031011}. 

In the single-photon limit, the influence of the temporal profile of the photon wavepacket on the atomic excitation has been studied in~\cite{Leong2016}. Furthermore, experimental work has been conducted that considers the scattering of two photons through 
a quantum-dot TLS, generating highly nonclassical fields such as photon bound states~\cite{Masters2023,Tomm2023}.

From a theoretical perspective, few-photon transport in waveguide QED systems has been extensively investigated using methods such as Green-function techniques, which are typically limited to only one or two photons, and usually rely on the weak excitation approximation (WEA), thus neglecting strong interactions which create finite populations; another common method is a master equation approach, in which the waveguide modes are traced out, removing any explicit field dynamics and strong-matter coupling effects (which can be very rich in a waveguide system), and missing important information such as photon-matter entanglement~\cite{Chang_2012,Mirhosseini2019}. 
While these are often effective and widely applied techniques, they rely on approximations that limit their use in certain cases, especially in waveguide-QED. Fortunately, there are a few powerful techniques that go beyond these approximations, such as matrix product states 
(MPS)~\cite{PhysRevLett.116.093601,Droenner2019,PhysRevResearch.3.023030,sofia2025}, scattering-matrix theories~\cite{PhysRevA.82.063821,PhysRevLett.98.153003,PhysRevA.76.062709,Rephaeli2012FewPhotonSC}, and analytically (or numerically) solving the relevant Heisenberg equations of motion~\cite{PhysRevA.83.063842,PhysRevA.106.023708,Nysteen2015,Chen_2011,PhysRevA.65.033832,PhysRevA.93.033838,PhysRevA.99.023857,PhysRevA.90.063832}. 

\blue{The temporal evolution of these quantum pulses determines important processes, including emitter excitation, pulse reshaping, field correlations, and the generation of nonclassical output states. Thus, access to the full time-dependent dynamics is crucial for linking theoretical predictions to time-resolved experiments and assessing the performance of quantum photonic devices. While stationary scattering approaches provide valuable information about input-output relations, they do not generally provide direct access to temporal light-matter dynamics during the interaction process.
}

In this paper, we directly compare two powerful approaches widely used to explore nonlinear interactions in waveguide QED, scattering matrix theory and MPS theory, but with a specific focus on the temporal dynamics with finite pulse durations from quantum fields. Although the scattering matrix approach has been broadly used to study few-photon scattering processes~\cite{PhysRevA.76.062709,PhysRevLett.98.153003}, the inclusion of different pulse envelopes is not well established in the literature, apart from a few works, such as Ref.~\cite{PhysRevLett.126.023603}. In addition, often scattering theories assume the long-time solution, disregarding the time-dynamics during the interaction. This long-time solution enables a powerful solution that can provide deep insight into the output field behavior, such as photon transmission and reflection, and field correlations; however, it contains a few limitations and challenges, including not being able to directly calculate the population dynamics of the TLS (though this can be possible in certain cases), and the difficulty of extending this method to higher photon numbers where it rapidly becomes intractable.
It is also not obvious how to include finite pulse excitation dynamics. Here, we will present a solution to extract the TLS population dynamics and other time-dependent correlation functions (at the few-photon level), and then directly show how this 
compares with another powerful approach that can overcome the finite photon 
restriction, based on MPS.

The MPS theory consists of a one-dimensional tensor network approach that discretizes the waveguide in time (time bins), in order to reduce the rapidly increasing Hilbert space~\cite{mcculloch_density-matrix_2007,orus_practical_2014}. This is an
extremely robust method capable of solving nonlinear waveguide QED problems that also involve retardation effects (time-delayed coherent feedback), light-matter entanglement, and a higher number of photons, with a significantly reduced numerical cost. Thus, it allows one to study, for example, Fock state pulses containing 
$N$-photons~\cite{PhysRevLett.116.093601,PhysRevA.103.033704}.

All the MPS calculations shown here have been performed using the open-source Python package QwaveMPS, which allows one to model waveguide QED systems using MPS~\cite{2602.15826}.

The rest of the paper is organized as follows. In Sec.~\ref{sec:SC_theory}, we introduce the scattering matrix theory, with a full derivation of this method when one considers finite pulses, including how to extract time-dependent populations (for the TLS and fields) and photon correlation functions. We start from a general Hamiltonian and consider the dynamics both when the TLS is symmetrically coupled to the waveguide, and when it is chirally coupled to only one direction. We first derive the single-photon case in Sec.~\ref{subsec:SC_single_ph}, and then we extend it to a two-photon pulse subspace in Sec.~\ref{subsec:SC_two_ph}, where in both cases we study transmission, reflection, and correlation dynamics. In Sec.~\ref{subsec:SC_pop}, we introduce a method to extract the TLS population dynamics from the photon properties and observables. 

We then present the MPS approach in Sec.~\ref{sec:MPS_theory}, where we give a summary of the theory and introduce an input pulse with a general shape containing $N$-photons written as a matrix product state, as well as the two-time first- and second-order photon correlation functions written as matrix product operators (MPOs).

Numerical results regarding both techniques are shown Sec.~\ref{sec:results},
including a comparison of calculations using scattering matrix theory and MPS for the first-order correlation functions in Sec.~\ref{subsec:g1}, and second-order correlations and populations in Sec.~\ref{subsec:g2}. We discuss and demonstrate some advantages of using scattering matrices in Sec.~\ref{subsec:pulse_length} with a study of the linear and nonlinear contributions for pulses of different lengths, and the full power of MPS is shown in Sec.~\ref{subsec:higher_photons}, where we calculate results for higher photon numbers (up to 8 photons).
Finally, in Sec.~\ref{sec:conclusions}, we present our conclusions.

\begin{figure}[t]
    \centering
    \includegraphics[width=\columnwidth]{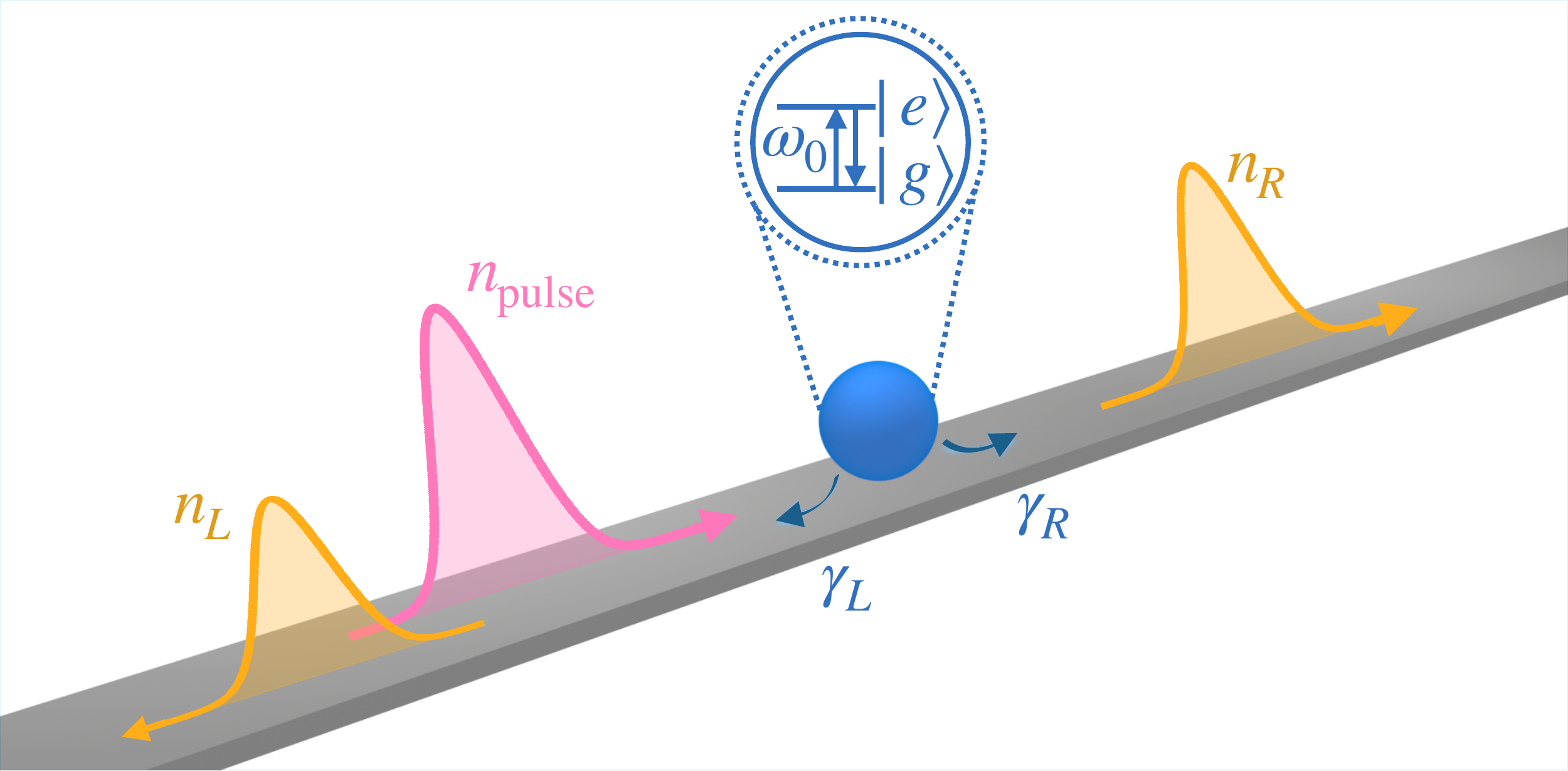}
    \caption{Schematic of the light-matter system of interest, which includes a TLS coupled to an infinite 
    (i.e., open) waveguide, where $\gamma_R$ and $\gamma_L$ correspond to the right/left coupling, and the emitter is excited by a quantum pulse
    (Fock state) containing one or a few photons with a pulse envelope shape $f(t)$. Here, $n_{\rm pulse}$ represents the flux of the input pulse, and $n_R$ and $n_L$ represent the right/left photon fluxes after the interaction with the TLS. For a chiral emitter, then  $\gamma_L=0$. }
    \label{fig:schematic}
\end{figure}

\section{Scattering Matrix Theory}
\label{sec:SC_theory}
We will first describe the main features of the scattering matrix approach, 
and also show how one can obtain time-dependent
observables~\cite{PhysRevLett.98.153003,PhysRevA.76.062709,PhysRevA.82.063821,PhysRevLett.126.023603,PhysRevLett.119.153601}.

The general expression of a single TLS 
coupled to a waveguide is defined by the following Hamiltonian,
\begin{equation}
    H = H_{\rm W} + H_{\rm TLS} + H_{\rm I},
\end{equation}
where
\begin{equation}
    H_{\rm W}= \sum_{\mu=\rm L,R}\int_{-\infty}^\infty d \omega \omega a^{\mu \ \dagger}(\omega) a^\mu(\omega),
\end{equation}
represents the waveguide modes, where the superscript $\mu = \rm L,R$ refer to the left and right channel, respectively. Note that the frequency integrations run from
$-\infty$ to $\infty$ (first Markov approximation~\cite{gardiner_zoller_2010}), as we will consider near-resonant interactions.

In the interaction picture, at the frequency 
$\omega_0$, the TLS Hamiltonian is
\begin{equation}
    H_{\rm TLS} = \frac{1}{2} \Delta \sigma^z ,
\end{equation}
where $\Delta=\omega_{\rm TLS}-\omega_0$ is the detuning between the TLS transition frequency ($\omega_{\rm TLS}$) and the center frequency of our chosen rotating frame ($\omega_0$), $\sigma^z = \sigma^+\sigma^--\sigma^-\sigma^+$, and $\sigma^{+/-}$ are the TLS Pauli matrices.
In a rotating wave approximation,
\begin{equation}
   H_{\rm I} = \sum_{\mu=\rm L,R}\sqrt{\frac{\gamma_\mu}{2\pi}} \int_{-\infty}^\infty d \omega \left( \sigma^+ a^\mu(\omega) + a^{\mu \ \dagger}(\omega) \sigma^- \right),
\end{equation}
represents the interaction between the TLS and the waveguide photons, with $\gamma_{\rm L}/\gamma_{\rm R}$ the left/right coupling rates.

We next define the input and output operators for photons~\cite{gardiner_zoller_2010}. The input operator is defined from,
\begin{equation}
    a^\mu_{\rm in}(t) = \frac{1}{\sqrt{2\pi}} \int_{-\infty}^\infty d \omega \ a^\mu (\omega,t_0) e^{-i\omega(t-t_0)},
\end{equation}
where $t_0 \ll t_{\rm int}$ is a previous time far from the interaction time $t_{\rm int}$. The output operator is
\begin{equation}
    a^\mu_{\rm out}(t) = \frac{1}{\sqrt{2\pi}} \int_{-\infty}^\infty d \omega \ a^\mu(\omega,t_1) e^{-i\omega(t-t_1)},
\end{equation}
where $t_1 \gg t_{\rm int}$, is a time much
greater than the interaction. The scattering theory is defined in the limit where $t_0 \to -\infty$ and $t_1 \to \infty$, and we can write:
\begin{equation}
    a^\mu_{\rm in}(t) = \frac{1}{\sqrt{2\pi}} \int_{-\infty}^\infty d \omega a^\mu_{\rm  in} (\omega) e^{-i\omega t},
    \label{ain}
\end{equation}
and
\begin{equation}
    a^\mu_{\rm out}(t) = \frac{1}{\sqrt{2\pi}} \int_{-\infty}^\infty d \omega a^\mu_{\rm out} (\omega) e^{-i\omega t}.
    \label{aout}
\end{equation}

The relevant equations of motion to obtain the input and output fields are
\begin{equation}
\begin{split}
    &\frac{d ( \sigma^+ (t) \sigma^-(t)) }{dt} = \sum_{\mu=\rm L,R} -i\sqrt{\gamma_\mu} \big[ \sigma^+ (t) a^\mu_{\rm in}(t) \\
    &   - a_{\rm in}^{\mu \ \dagger}(t) \sigma^- (t)\big] - \gamma  \sigma^+ (t)\sigma^-(t),
    \label{heis1}
\end{split}
\end{equation}
\begin{equation}
\begin{split}
    \frac{d \sigma^-(t)}{dt} 
    &= \sum_{\mu=\rm L,R}i\sqrt{\gamma_\mu}\sigma^z(t) a^\mu_{\rm in}(t) \\
    &- \frac{\gamma}{2}\sigma^-(t) - i\Delta \sigma^-(t),
    \label{heis2}
\end{split}
\end{equation}
and
\begin{equation}
    a^\mu_{\rm out}(t) = a^\mu_{\rm in}(t) - i\sqrt{\gamma_\mu}\sigma^-(t),
    \label{heis3}
\end{equation}
where $\gamma = \gamma_{\rm R} +\gamma_{\rm L}$ is the total coupling rate. 

In a chiral emitter (`ch') system, we have only one channel scattering, and we can rewrite the field operators $a^\mu(t) = a^{\rm ch}(t)$ and the single-channel decay rate $\gamma=\gamma_{\rm R}$. Thus, the equations of motion simplify to,
\begin{equation}
\begin{split}
    &\frac{d ( \sigma^+ (t) \sigma^-(t)) }{dt} = \\
    &-i\sqrt{\gamma} \left[ \sigma^+ (t) a^{\rm ch}_{\rm in}(t) - a_{\rm in}^{\rm ch \ \dagger} (t) \sigma^- (t)\right] - \gamma  \sigma^+ (t)\sigma^-(t),
    \label{heis1_ch}
\end{split}
\end{equation}
\begin{equation}
    \frac{d \sigma^-(t)}{dt} = i\sqrt{\gamma}\sigma^z(t) a^{\rm ch}_{\rm in}(t) - \frac{\gamma}{2}\sigma^-(t) - i\Delta \sigma^-(t),
    \label{heis2_ch}
\end{equation}
and
\begin{equation}
    a^{\rm ch}_{\rm out}(t) = a^{\rm ch}_{\rm in}(t) - i\sqrt{\gamma}\sigma^-(t).
    \label{heis3_ch}
\end{equation}

Note that we obtain the expected spontaneous emission decay in the absence of any excitation [Eqs.~\eqref{heis1} and~\eqref{heis1_ch}], which is an important distinction compared to the harmonic oscillator case. The scattering matrices for different cases can be found by using these Heisenberg equations of motion.

\subsection{Single photon scattering}
\label{subsec:SC_single_ph}

An initial state, containing a single photon propagating to the right, can be defined as~\cite{PhysRevLett.126.023603}
\begin{equation}
    \ket{\psi_{\rm in}^{(1)}} = \int d\omega f(\omega) a_{\rm in}^{\rm{R} \ \dagger}(\omega) \ket{0},
\end{equation}
where $f(\omega)$ represents the wavepacket of the single photon in frequency space, and
$\int_{-\infty}^{\infty} d\omega |f(\omega)|^2=1$.
This is related via the Fourier transform to the temporal waveform of the single photon in the time domain; we stress this connection, as often scattering theory is used in a CW picture
(single frequency input), where
$|f(\omega)|^2= \delta(\omega-\omega_p)$.
In the time domain, such a limit does not make much sense, since it implies that
$f(t) \rightarrow 0$ for the single photon amplitude. In practice, 
single photons, for example, from a classically pulse-triggered quantum emitter,
must be at least photons per unit time. So generally, one must deal with pulsed
single photons and pulsed few-photons coming into the waveguide QED system.

To connect the frequency-dependent photon operators to time-dependent ones,
using Eqs.~\eqref{ain} and~\eqref{aout}, we define
\begin{equation}
    a_{\rm in/out}^\mu(\omega) = \frac{1}{\sqrt{2\pi}} \int_{-\infty}^\infty d t a_{\rm in/out}^\mu (t) e^{i\omega t}.
\end{equation}
With this, one can now calculate the single-photon scattering matrix, 
\begin{equation}
\begin{split}
    S^\mu_{\nu \omega} &= \braket{1_\nu^\mu |1_\omega^{\rm R}}= \bra{g,0} a^\mu_{\rm out}(\nu)a_{\rm in}^{\rm R\dag}(\omega) \ket{0,g} \\
    &= \bra{0} a^{\mu}_{\rm out}(\nu) \ket{1_\omega^\mu},
    \label{S1ph}
\end{split}
\end{equation}
where $\omega$ is the input field frequency (which now covers a bandwidth), and $\nu$ is the output one (also with bandwidth, in the case of a temporal input pulse).
It is important to note that we are considering the TLS initially in the ground state, i.e., $\ket{g}$.
However, this does not need to be the case,
though relaxation of this assumption may considerably complicate the scattering theory solutions.

Using Eq.~\eqref{heis3_ch}, the chiral case, where there is a single-photon channel, can be calculated from
\begin{equation}
    \bra{0} a^{\rm ch}_{\rm out} (t) \ket{1_\omega} = \bra{0} a^{\rm ch}_{\rm in} (t)\ket{1_\omega} - i \sqrt{\gamma}\bra{0} \sigma^- (t)\ket{1_\omega}
    \label{aout_t}
\end{equation}
with
\begin{equation}
\begin{split}
    \bra{0} a_{\rm in} (t) \ket{1_\omega} &= \frac{1}{\sqrt{2\pi}} \int d \omega' \bra{0} a_{\rm in}(\omega') e^{-i\omega' t} a^{ \dagger}_{\rm in}(\omega)\ket{0} \\ 
    &= \frac{1}{\sqrt{2\pi}}  e^{-i\omega t}.
\end{split}
\end{equation}

From Eq.~\eqref{heis2_ch}, we derive
\begin{equation}
\begin{split}
    &\frac{d}{dt} \bra{0} \sigma^-(t) \ket{1_\omega} = i \sqrt{\gamma} \bra{0} \sigma^z(t) \ket{1_\omega} \\
    &- \left( \frac{\gamma}{2} + i\Delta \right) \bra{0} \sigma^-(t) \ket{1_\omega} \\
    &= -i\sqrt{\frac{\gamma}{2\pi}}e^{-i\omega t } -  \left( \frac{\gamma}{2} + i\Delta \right) \bra{0} \sigma^-(t) \ket{1_\omega},
    \label{d_sigma_1ph}
\end{split}
\end{equation}
where we have replaced $\sigma^z \rightarrow -\mathbb{1}$ 
in the one-photon limit. Integrating Eq.~\eqref{d_sigma_1ph}, then
\begin{equation}
    \bra{0} \sigma^- (t) \ket{1_\omega} = \frac{1}{\sqrt{2\pi}} \frac{\sqrt{\gamma}}{i\gamma/2 - \Delta + \omega} e^{-i\omega t},
\end{equation}
and introducing these in Eq.~\eqref{aout_t}:
\begin{equation}
    \bra{0} a^{\rm ch}_{\rm out} (t) \ket{1_\omega} = \frac{1}{\sqrt{2\pi}} \frac{-i\gamma/2 - \Delta + \omega}{i\gamma/2 - \Delta +\omega} e^{-i\omega t}.
\end{equation}

In the frequency domain, we obtain
\begin{equation}
\begin{split}
    S_{\nu \omega} &= \bra{0} a^{\rm ch}_{\rm out} (\nu) \ket{1_\omega} = 
    \frac{-i\gamma/2 - \Delta + \omega}{i\gamma/2 - \Delta +\omega} \delta(\nu-\omega) \\ &= \chi(\omega)  \delta(\nu-\omega),
\end{split}
\end{equation}
where we have introduced the chiral transmission coefficient,
\begin{equation}
    \chi(\omega) \equiv t_{\rm ch}(\omega) = \frac{-i\gamma/2 - \Delta + \omega}{i\gamma/2 - \Delta +\omega}. 
\end{equation}

We can now extend this approach to the symmetrical emitter case, where the scattering matrix is now:
\begin{equation}
    S^\mu_{\nu \omega} = \chi^\mu (\omega)  \delta(\nu-\omega),
    \label{eq:1phS}
\end{equation}
with $\chi^{\mu}=\chi^{\rm R,L}=t_{\rm sym},r_{\rm sym}$ the transmission (transmitted to the right) and reflection (reflected to the left) coefficients, respectively. In the resonant case between the TLS transition and the pulse center frequency, where $\Delta=0$~\cite{PhysRevLett.126.023603}, we can write the general solution as
\begin{equation}
    t_{\rm sym}(\omega) = 1+r_{\rm sym}(\omega),
\end{equation}
with the reflection coefficient
\begin{equation}
    r_{\rm sym}(\omega) = \frac{-\gamma}{\gamma -2i(\omega-\omega_0)}.
\end{equation}

\subsubsection{Output state}

Next, we calculate the output state in the single photon case as a superposition of single-photon Fock states, which can either be transmitted or reflected, with a wavepacket $f(\omega)$. Using the result from Eq.~\eqref{eq:1phS},  then
\begin{equation}
\begin{split}
        \ket{\psi_{\rm out}^{(1)}} &= \sum_{\mu=L,R} \iint d\nu d\omega f(\omega) S_{\nu \omega}^\mu  \ket{1_\nu^\mu} \\
        &=  \sum_{\mu=L,R} \int d\omega f(\omega) \chi^\mu (\omega)\ket{1_\omega^\mu}, 
\end{split}
\end{equation}
and the projected wavefunction on $\mu$, is
\begin{equation}
\begin{split}
    \varphi_\mu^{(1)}(t) &= \frac{1}{\sqrt{2\pi}} \int d\omega e^{-i\omega t} \bra{0} a_{\rm in}^\mu (\omega) \ket{\psi_{\rm out}^{(1)}}  \\
    &= \frac{1}{\sqrt{2\pi}} \int d\omega e^{-i \omega t}  f(\omega) \chi^\mu (\omega).
    \end{split}
    \label{1ph_proj}
\end{equation}
Note that we drop the explicit integral limits for simpler notation, but recall that the frequency integrations run from $-\infty$ to $\infty$.

\subsubsection{First-order two-time correlation function}

In the single-photon subspace, the first-order quantum correlation function, applied at two different times, is defined as
\begin{widetext}
\begin{equation}
\begin{split}
    &G^{(1)}_{1,\mu\mu'}(t,t+\tau) = \braket{a_{\rm out}^{\mu \ \dagger} (t)a_{\rm out}^{\mu'}(t+\tau) } 
    = \bra{\psi_{\rm in}^{(1)}}  a_{\rm out}^{\mu \ \dagger} (t)a_{\rm out}^{\mu'}(t+\tau) \ket{\psi_{\rm in}^{(1)}} \\
    &= \iint d\omega d\nu \bra{0} f^*(\nu) a_{\rm in}^{\rm R}(\nu) a_{\rm out}^{\mu \ \dagger} (t)a_{\rm out}^{\mu'}(t+\tau) f(\omega) a_{\rm in}^{\rm R \ \dagger}(\omega) \ket{0} \\
    &=\frac{1}{2\pi} \iint d\omega d\nu \iint dk dp f^*(\nu)e^{ik t}e^{-ip(t+\tau)} f(\omega)\braket{1_\nu^{\rm R}| a_{\rm out}^{\mu \ \dagger} (k)|0}\braket{0| a_{\rm out}^{\mu'}(p) |1_\omega^{\rm R}} \\
    &= \frac{1}{2\pi} \iint d\omega  d\nu f^*(\nu) \chi^{\mu *} (\nu) e^{i\nu t} \chi^{\mu '}(\omega) e^{-i\omega(t+\tau)} f(\omega) =  \varphi_\mu^{(1) *}(t)  \varphi_{\mu'}^{(1)}(t+\tau),
\end{split}
\end{equation}
\end{widetext}
where we have used the identity operator, as well as Eqs.~\eqref{S1ph} and~\eqref{1ph_proj}.

\subsection{Two photon scattering}
\label{subsec:SC_two_ph}

The input state for the two-photon transport case can be written as follows: 
\begin{equation}
    \ket{\psi_{\rm in}^{(2)}} = \iint \frac{d \omega_1 d\omega_2}{\sqrt{2}} f(\omega_1,\omega_2) a_{\rm in}^{\rm R  \ \dagger}(\omega_1)a_{\rm in}^{\rm R  \ \dagger}(\omega_2) \ket{0} ,
\end{equation}
where we consider the input photons traveling to the right; 
the factor of $\frac{1}{\sqrt{2}}$ comes from the need for the input state to be normalized, with $\iint d\omega_1 d\omega_2 |f(\omega_1,\omega_2)|^2 = 1$, and  $f(\omega_1,\omega_2) = f(\omega_2,\omega_1)$ is symmetrical under photon exchange. 

The scattering matrix for two photons can be derived in a way similar to the one-photon solution, where 
\begin{equation}
\begin{split}
    &S^{\mu,\mu'}_{\nu_1\nu_2\omega_1\omega_2} = 
    \braket{1_{\nu_1}^{\mu} 1_{\nu_2}^{\mu'} |1_{\omega_1}^{\rm R} 1_{\omega_2}^{\rm R}} \\
    &=\bra{0}a_{\rm out}^{\mu} (\nu_1) a_{\rm out}^{\mu'} (\nu_2) a_{\rm in}^{\rm R \ \dagger} (\omega_1) a_{\rm in}^{\rm R \ \dagger} (\omega_2)\ket{0}. 
\end{split}
\end{equation}

The scattering matrix for the symmetrical case is now~\cite{PhysRevA.82.063821,PhysRevLett.126.023603}
\begin{equation}
\begin{split}
    S^{\mu,\mu'}_{\nu_1\nu_2\omega_1\omega_2} &= \chi^\mu (\nu_1) \chi^{\mu '} (\nu_2) \big[\delta(\nu_1-\omega_1) \delta(\nu_2-\omega_2) \\
    &+ \delta(\nu_1-\omega_2) \delta(\nu_2-\omega_1) \big] \\
    &+ T^{\rm sym}_{\nu_1\nu_2\omega_1\omega_2} \delta(\nu_1+\nu_2-\omega_1-\omega_2),
\end{split}
\end{equation}
with 
\begin{equation}
    T^{\rm sym}_{\nu_1\nu_2\omega_1\omega_2} = \frac{4}{\pi \gamma} \frac{r_{\rm sym}(\nu_1)r_{\rm sym}(\nu_2) r_{\rm sym}(\omega_1)r_{\rm sym}(\omega_2)}{r_{\rm sym}\left(\frac{\omega_1+\omega_2}{2}\right)}.
    \label{T_sym}
\end{equation}
Note that all the transmission and reflection coefficients here are obtained from the one-photon solution equations,
and they represent linear response functions.  

In the case of a chiral emitter system~\cite{PhysRevA.82.063821}, then
\begin{equation}
\begin{split}   
S_{\nu_1\nu_2\omega_1\omega_2} &= t_{\rm ch}(\nu_1)t_{\rm ch}(\nu_2) \big[\delta(\nu_1 - \omega_1) \delta(\nu_2-\omega_2) \\
&+ \delta(\nu_1-\omega_2) \delta(\nu_2-\omega_1) \big] \\
    &+ T^{\rm ch}_{\nu_1\nu_2\omega_1\omega_2} \delta( \nu_1 + \nu_2 - \omega_1-\omega_2) , 
\end{split}    
\end{equation}
where in this case,
\begin{equation}
    T^{\rm ch}_{\nu_1\nu_2\omega_1\omega_2} =  \frac{i\sqrt{\gamma}}{\pi} s(\nu_1) s(\nu_2) \left[  s(\omega_1) + s(\omega_2) \right], 
\end{equation}
with 
\begin{equation}
s(\omega) = \frac{\sqrt{\gamma}}{(\omega-\Delta) + i\gamma/2},
\end{equation}
where $s$ has units of $\sqrt{t}$, and this solution uses the calculation $\bra{0} \sigma^- \ket{1_\omega}$ in the single-photon subspace. 

\subsubsection{Output state}

Using the completeness relation,
$1/2 \int dp_1 \int d p_2 \ket{p_1 p_2} \bra{p_1 p_2} =1$, the output state can now be written as 
\begin{equation}
\begin{split}
    &\ket{\psi_{out}^{(2)}} = \sum_{\mu,\mu'\in\{\rm L,R\}}\iint \frac{d\nu_1 d\nu_2}{2\sqrt{2}} \\
    & \iint d \omega_1 d\omega_2 f(\nu_1,\nu_2)S^{\mu,\mu'}_{\nu_1\nu_2\omega_1\omega_2} a_{\rm in}^{\mu  \ \dagger}(\omega_1)a_{\rm in}^{\mu'  \ \dagger}(\omega_2) \ket{0}.
\end{split}
\end{equation}

In the symmetrical coupling case,
\begin{equation}
\begin{split}
    &\ket{\psi_{out}^{(2)}}
    = \frac{1}{\sqrt{2}} \sum_{\mu,\mu'}\iint d\omega_1 d\omega_2 \big\{ f(\omega_1,\omega_2)  \chi^\mu (\omega_1) \chi^{\mu '} (\omega_2) \\
    &+ \frac{1}{2}\mathcal{I}^{\rm sym}_{\omega_1 \omega_2} \big\} a_{\rm in}^{\mu  \ \dagger}(\omega_1)a_{\rm in}^{\mu'  \ \dagger}(\omega_2) \ket{0},
    \label{psi2}
\end{split}
\end{equation}
where 
\begin{equation}
    \mathcal{I}^{\rm sym}_{\omega_1 \omega_2} = \iint d\nu_1 d\nu_2 f(\nu_1,\nu_2)T^{\rm sym}_{\nu_1\nu_2\omega_1\omega_2} \delta(\nu_1+\nu_2-\omega_1-\omega_2).
\end{equation}
To solve this term, we can use the following change of variables: 
\begin{equation}
\begin{split}
    &\omega' = \frac{\nu_1 + \nu_2}{2}; \ \ \Delta'=\frac{\nu_2 - \nu_1}{2} \\
    &\Rightarrow \nu_1 = \omega' -\Delta' ; \ \ \nu_2 = \omega'+ \Delta'; \ \ d\nu_1 d\nu_2 = 2d\omega' d\Delta',
\end{split}
\end{equation}
and thus, 
\begin{equation}
\begin{split}
    \mathcal{I}^{\rm sym}_{\omega_1 \omega_2} 
    &= \int d \Delta' f(\omega' - \Delta',\omega' + \Delta') T^{\rm sym}_{\omega' - \Delta', \omega' +\Delta',\omega_1,\omega_2 }.
\end{split}
\end{equation}

We can also calculate the projected wavefunction of two photons, now on $t,t+\tau$ and $\mu,\mu'$:
\begin{widetext}
\begin{equation}
\begin{split}
    &\varphi_{\mu,\mu'}^{(2)}(t,t+\tau) = \iint \frac{d\omega_1 d\omega_2}{2\pi}\bra{2_{\omega_1\omega_2}^{\mu\mu'}}e^{-i\omega_1 t} e^{-i\omega_2(t+\tau)}\ket{\psi_{out}^{(2)}} = \iint \frac{d\omega_1 d\omega_2}{2\sqrt{2} \pi}\bra{0} a_{\rm in}^\mu (\omega_1) a_{\rm in}^{\mu'} (\omega_2)e^{-i\omega_1 t} e^{-i\omega_2(t+\tau)}  \times\\
& \sum_{\eta,\eta'\in\{\rm L,R\}}\iint d\nu_1 d\nu_2 \big\{f(\nu_1,\nu_2) 
 \chi^\eta (\nu_1) \chi^{\eta '} (\nu_2) + \frac{1}{2}\int d \Delta' f(\omega'-\Delta',\omega'+\Delta') T^{\rm sym}_{\omega' - \Delta', \omega' +\Delta',\nu_1,\nu_2 } \big\} a_{\rm in}^{\eta  \ \dagger}(\nu_1)a_{\rm in}^{\eta'  \ \dagger}(\nu_2) \ket{0} \\
&=  \iint \frac{d\omega_1 d\omega_2}{\sqrt{2}\pi}e^{-i\omega_1 t} e^{-i\omega_2(t+\tau)} \big\{f(\omega_1,\omega_2) \chi^\mu (\omega_1) \chi^{\mu '} (\omega_2) 
+ \frac{1}{2}\int d \Delta' f(\omega'-\Delta',\omega'+\Delta')T^{\rm sym}_{\omega' - \Delta', \omega' +\Delta',\omega_1,\omega_2 } \big\} \\
&=  \iint \frac{d\omega_1 d\omega_2}{\sqrt{2}\pi}e^{-i\omega_1 t} e^{-i\omega_2(t+\tau)} \big\{ I^{\mu,\mu'}_{\rm lin} + I^{\rm sym}_{\rm nlin} \big\}, 
\end{split}
\label{eq:2ph_proj}
\end{equation}
\end{widetext}
where we have defined 
\begin{equation}
    I^{\mu,\mu'}_{\rm lin} = f(\omega_1,\omega_2) \chi^\mu (\omega_1) \chi^{\mu '} (\omega_2),
    \label{Ilin}
\end{equation}
and
\begin{equation}
    I^{\rm sym}_{\rm nlin} =
\frac{1}{2}\int d \Delta' f(\omega'-\Delta',\omega'+\Delta')T^{\rm sym}_{\omega' - \Delta', \omega' +\Delta',\omega_1,\omega_2 }. 
\label{Inlin}
\end{equation}

The same procedure is then used for a chiral system, where the output state is
\begin{equation}
\begin{split}
    &\ket{\psi_{\rm out}^{(2)}}  = \frac{1}{\sqrt{2}} \iint d\omega_1 d\omega_2 \big\{ f(\omega_1,\omega_2)  t_{\rm ch} (\omega_1) t_{\rm ch} (\omega_2) \\
    &+ \frac{1}{2}\mathcal{I}^{\rm ch}_{\omega_1 \omega_2} \big\} a_{\rm in}^{\rm ch \ \dagger}(\omega_1)a_{\rm in}^{\rm ch \ \dagger}(\omega_2) \ket{0},
\end{split}
\end{equation}
with 
\begin{equation}
    \mathcal{I}^{\rm ch}_{\omega_1 \omega_2} = \iint d\nu_1 d\nu_2 f(\nu_1,\nu_2)T^{\rm ch}_{\nu_1\nu_2\omega_1\omega_2} \delta(\nu_1+\nu_2-\omega_1-\omega_2),    
\end{equation}
and the projection:
\begin{equation}
\begin{split}
     &\varphi_{\rm ch}^{(2)}(t,t+\tau) =
      \iint \frac{d\omega_1 d\omega_2}{\sqrt{2}\pi}e^{-i\omega_1 t} e^{-i\omega_2(t+\tau)}  \\ 
      & \times\big\{f(\omega_1,\omega_2) t_{\rm ch}(\omega_1) t_{\rm ch}(\omega_2) \\
    &+ \frac{1}{2}\int d \Delta' f(\omega'-\Delta',\omega'+\Delta')T^{\rm ch}_{\omega' - \Delta', \omega' +\Delta',\omega_1,\omega_2 } \big\} \\
     &= 
     \frac{1}{\sqrt{2}\pi} \iint d\omega_1 d\omega_2 e^{-i\omega_1 t} e^{-i\omega_2(t+\tau)}\big\{
     I^{\rm ch}_{\rm lin} + I^{\rm ch}_{\rm nlin} \big\}.
\end{split}
\end{equation}
\vspace{1.8em}
In this case, we have
\begin{equation}
    I^{\rm ch}_{\rm lin} = f(\omega_1,\omega_2) t_{\rm ch} (\omega_1) t_{\rm ch} (\omega_2),
\end{equation}
and
\begin{equation}
    I^{\rm ch}_{\rm nlin} =
\frac{1}{2}\int d \Delta' f(\omega'-\Delta',\omega'+\Delta')T^{\rm ch}_{\omega' - \Delta', \omega' +\Delta',\omega_1,\omega_2 }, 
\end{equation}
for the explicit linear and nonlinear terms.

\subsubsection{First-order two-time correlation function}

We next evaluate the first-order correlation function at two different times,

\begin{widetext}
\begin{equation}
\begin{split}
    &G_{2,\mu\mu'}^{(1)}(t,t+\tau) =\braket{\psi_{\rm{\rm in}}^{(2)}| a_{\rm{out}}^{\mu\ \dagger}(t)a_{\rm{out}}^{\mu'}(t+\tau) |\psi_{\rm{\rm in}}^{(2)}} \
    = \frac{1}{2\pi}\iint d\omega_1 d\omega_2 e^{i\omega_1 t}e^{-i\omega_2 (t+\tau)} \braket{\psi_{\rm{\rm in}}^{(2)}| a_{\rm{out}}^{\mu\ \dagger}(\omega_1)a_{\rm{out}}^{\mu'}(\omega_2) |\psi_{\rm{\rm in}}^{(2)}} \\
    =& \frac{1}{2\pi}\iint d\omega_1 d\omega_2 e^{i\omega_1 t}e^{-i\omega_2 (t+\tau)} \braket{\psi_{\rm{out}}^{(2)}| a_{\rm{\rm in}}^{\mu\ \dagger}(\omega_1) a_{\rm{\rm in}}^{\mu'}(\omega_2) |\psi_{\rm{out}}^{(2)}},
    \label{eq:G1_muMuP_Initial}
\end{split}    
\end{equation}
\end{widetext}
where we have Fourier transformed the operators to frequency space, and used the unitarity of the scattering matrix. Making use of the single photon subspace identity operator, we also have
\begin{equation}
\begin{split}
    &\braket{\psi_{\rm{out}}^{(2)}| a_{\rm{\rm in}}^{\mu\ \dagger}(\omega_1) a_{\rm{\rm in}}^{\mu'}(\omega_2) |\psi_{\rm{out}}^{(2)}} \\
    &= \sum_{\eta\in\{\rm L,R\}} \int d\nu \braket{\psi_{\rm{out}}^{(2)}| a_{\rm{\rm in}}^{\mu\ \dagger}(\omega_1)|1_\nu^\eta}\braket{1_\nu^\eta|  a_{\rm{\rm in}}^{\mu'}(\omega_2) |\psi_{\rm{out}}^{(2)}}. 
\end{split}    
\end{equation}

We define a function
\begin{widetext}
\begin{equation}
\begin{split}
    &\varphi_{2,\eta\mu}^{\rm{G1}}(t,\nu)\equiv \frac{1}{\sqrt{2\pi}}\int d\omega e^{-i\omega t}\braket{0| a_{\rm{\rm in}}^{\eta}(\nu)a_{\rm{\rm in}}^{\mu}(\omega) |\psi_{\rm{out}}^{(2)}} \\
    &= \frac{1}{\sqrt{2\pi}}\int d\omega e^{-i\omega t} \sum_{\lambda,\lambda'\in\{\rm L,R\}}\iint dp_1 dp_2 \frac{1}{\sqrt{2}} \big\{f(p_1,p_2) \chi^\lambda (p_1) \chi^{\lambda '} (p_2) + \frac{1}{2}\mathcal{I}_{p_1p_2} \big\} \braket{0| a_{\rm{\rm in}}^{\eta}(\nu) a_{\rm{\rm in}}^{\mu}(\omega)  a_{\rm in}^{\lambda  \ \dagger}(p_1)a_{\rm in}^{\lambda'  \ \dagger}(p_2)|0} \\
    &= \frac{1}{\sqrt{\pi}}\int d\omega e^{-i\omega t} \big\{ f(\omega,\nu)  \chi^\mu (\omega) \chi^{\eta} (\nu) + \frac{1}{2}\mathcal{I}^{\rm sym}_{\omega\nu} \big\},
\end{split}    
\end{equation}
\end{widetext}
and thus, we can express the first-order correlation as
\begin{align}
     G_{2,\mu\mu'}^{(1)}(t,t+\tau) =& \sum_{\eta\in\{\rm L,R\}} \int d\nu \varphi_{2,\eta\mu}^{\rm{G1}}(t,\nu)^* \varphi_{2,\eta\mu'}^{\rm{G1}}(t+\tau,\nu),
     \label{G1_2ph}
\end{align}
where we have used $\mathcal{I}_{\omega\nu} = \mathcal{I}_{\nu\omega}$.

In the chiral TLS scattering case, our Hilbert space is restricted to photons only in transmission. As a result, the chiral first-order correlation function is,
\begin{equation}
     G_{2, \rm ch}^{(1)}(t,t+\tau) = \int d\nu \varphi_{2, \rm ch}^{\rm{G1}}(t,\nu)^* \varphi_{2, \rm ch}^{\rm{G1}}(t+\tau,\nu),
\end{equation}
where 
\begin{equation}
    \varphi_{2, \rm ch}^{\rm{G1}}(t,\nu) = \frac{1}{\sqrt{\pi}}\int \!d\omega e^{-i\omega t} 
    \!
    \left\{ f(\omega,\nu)  t_{\rm ch} (\omega) t_{\rm ch}(\nu) + \frac{1}{2}\mathcal{I}^{\rm ch}_{\omega\nu} \right\}.
\end{equation}

\subsubsection{Second-order two-time correlation function}

Finally, to calculate the second-order correlation function, we  use~\cite{PhysRevLett.126.023603}
\begin{equation}
    G^{(2)}_{\mu, \mu'}(t,t+\tau) = |\varphi_{\mu,\mu'}^{(2)}(t,t+\tau)|^2,
    \label{G2_2ph}
\end{equation}
where $\varphi_{\mu,\mu'}^{(2)}(t,t+\tau)$
is obtained from Eq.~\eqref{eq:2ph_proj}. 
In the chiral case, this simplifies to,
\begin{equation}
    G_{\rm ch}^{(2)}(t,t+\tau) = |\varphi_{\rm ch}^{(2)}(t,t+\tau)|^2.
\end{equation}

\subsection{Population dynamics}
\label{subsec:SC_pop}

In a system with no losses, such that the total quanta in the system ($N_{\rm total}$) is conserved, the Hamiltonian commutes with the following operator,
\begin{equation}
    N = \sigma^+\sigma^- + N_{\rm wg},
\end{equation}
where
\begin{equation}
    N_{\rm wg} = \sum_{\mu\in\{L,R\}}\int_{-\infty}^\infty d\omega a^{\mu}(\omega)^\dagger a^{\mu}(\omega)
\end{equation}
is the photon population in the waveguide.
This can be used to compute the population of the two-level system as a function of time, $n_{\rm{TLS}} (t)\equiv \braket{\sigma^+\sigma^- (t)}$, which can be obtained from
\begin{align}
    n_{\rm{TLS}}(t) = N_{\rm total} - N_{\rm wg}(t).
    \label{eq:tls_pop}
\end{align}

The waveguide population is given by the time-dependent sum of the right and left channel photon populations [$n_{\rm R}(t)$ and $n_{\rm L}(t)$] integrated in time, and the population in the incident pulse remaining in the waveguide, which in the case of a TLS initially in the ground state is equivalent to the total number of excitations minus the time dependent photon flux into the TLS ($n_{\rm pulse}$),
\begin{align}
    N_{\rm wg}(t) &= \int_{-\infty}^t dt' n_{\rm R}(t') + \int_{-\infty}^t dt' n_{\rm L}(t') \nonumber \\ 
    &+ \left(N_{\rm total}-\int_{-\infty}^t dt' n_{\rm pulse}(t')\right)   \nonumber\\ 
    &= 
    N_{\rm R}(t) + N_{\rm L}(t) +  N_{\rm inc}(t),
\end{align}
where $N_{\rm R}(t)$ and $N_{\rm L}(t)$ correspond to the time-integrated right and left photon output fluxes, respectively, and  $N_{\rm inc}(t)$ is the integrated incident photon population before the interaction.

It is important to note here that the photon populations are equivalent to the corresponding first-order correlation functions applied at a single time:
\begin{equation}
n_{\rm R}(t) \equiv G_{\rm RR}^{(1)}(t,t), \ \  n_{\rm L} (t) \equiv G_{\rm LL}^{(1)}(t,t).
\end{equation}
In addition, the input pulse flux can be calculated as,
\begin{equation}
\begin{split}
    &n^{(m)}_{\rm pulse}(t)= \bra{\psi_{\rm in}^{(m)}} a^{\rm \mu \ \dagger}_{\rm in}(t) a^{\rm \mu}_{\rm in}(t) \ket{\psi_{\rm in}^{(m)}},
\end{split}    
\end{equation}
with $\psi_{\rm in}^{(m)}$ the initial state containing $m$ photons, and $\mu$ in this case corresponds to the same direction as the initial pulse.

In the cases studied here with 1-photon and 2-photon Fock states, respectively, then
\begin{widetext}
\begin{flalign}
    &n^{(1)}_{\rm pulse}(t)= \bra{\psi_{\rm in}^{(1)}} a_{\rm in}^{\rm \mu \ \dagger}(t)a^{\rm \mu}_{\rm in}(t) \ket{\psi_{\rm in}^{(1)}} & \nonumber \\
    &=\frac{1}{2\pi}\iint dk dp \ e^{ikt} e^{-ipt} \bra{0} \int d\nu f^*(\nu) a_{\rm in}^{\rm{\mu} }(\nu) a_{\rm in}^{\rm \mu \ \dagger}(k)a^{\rm \mu}_{\rm in}(p)\int d\omega f(\omega) a_{\rm in}^{\rm{\mu} \ \dagger}(\omega) \ket{0} \nonumber \\
    &= \frac{1}{2\pi}\iint dk dp \ e^{ikt} e^{-ipt} \bra{0} f^*(k) f(p) \ket{0}   = \left| f(t) \right|^2, 
\end{flalign}
%
and
\begin{equation}
\begin{split}
    &n^{(2)}_{\rm pulse}(t)= \bra{\psi_{\rm in}^{(2)}} a_{\rm in}^{\rm \mu \ \dagger}(t)a^{\rm \mu}_{\rm in}(t) \ket{\psi_{\rm in}^{(2)}} \\
    &=\frac{1}{2\pi}\iint dk dp \ e^{ikt} e^{-ipt} \bra{0} \iint \frac{d \nu_1 d\nu_2}{\sqrt{2}} f^*(\nu_1,\nu_2) a_{\rm in}^{\rm \mu }(\nu_1)a_{\rm in}^{\rm \mu}(\nu_2) a_{\rm in}^{\rm \mu \ \dagger}(k)a^{\rm \mu}_{\rm in}(p)\iint \frac{d \omega_1 d\omega_2}{\sqrt{2}} f(\omega_1,\omega_2) a_{\rm in}^{\rm \mu  \ \dagger}(\omega_1)a_{\rm in}^{\rm \mu  \ \dagger}(\omega_2) \ket{0} \\
    &= \frac{2}{2\pi}\iint dk dp \ e^{ikt} e^{-ipt} \int d\omega  f^*(\omega,k) f(\omega,p)
    = 2 \int d \omega \left| f(\omega,t) \right|^2,
\end{split}    
\end{equation}
\end{widetext}
where we have used the assumption that the two-photon pulse envelope is symmetric, hence, $f(\omega_1,\omega_2) = f(\omega_2,\omega_1)$. 

Finally, using Eq~\eqref{eq:tls_pop}, we can calculate the TLS population,
\begin{equation}
\begin{split}
    n_{\rm{TLS}}(t) = \int_{-\infty}^t dt' n^{(m)}_{\rm pulse}(t') - N_{\rm R}(t) - N_{\rm L}(t).
    \label{eq:tls_pop_f}
\end{split}    
\end{equation}

\section{Matrix Product States}
\label{sec:MPS_theory}

Next, we describe the MPS approach to model the waveguide-QED systems.
In the MPS scheme, tensor networks are used to describe the system, including the TLS and the waveguide, with the latter discretized in time bins in the Fock state basis~\cite{PhysRevLett.116.093601}. 

With this approach, a one-photon pulse can be introduced from~\cite{Guimond_2017,PhysRevA.103.033704,sofia2025}

\blue{
\begin{equation}
        \ket{\phi_0^{(1)}}= b_{\rm in}^\dagger \ket{0,...,0}, 
        \label{eq:phi_0}
\end{equation}
where $\ket{0,...,0}$ with each 0 corresponding to the vacuum state of a discretized portion of the waveguide represents the waveguide in vacuum, and the superscript `1' relates to $N=1$. The input pulse is created using,
\begin{equation}
    b_{\rm in}^\dagger = \int dt \ f(t) b_\mu^\dagger(t),
    \label{bin}
\end{equation}
where $\mu = \rm L,R$ correspond to the left or right channel, meaning that the initial pulse can propagate in one of these directions with the normalized pulse envelope $f(t)$, and the quantum noise creation operator in the time domain,}
\begin{equation}
    b^\dagger_\mu(t)=\frac{1}{\sqrt{2\pi}} \int d\omega b^\dagger_\mu(\omega) e^{i(\omega - \omega_{0})t},
    \label{continuous-time-ops}
\end{equation}
which follows the commutation relations $\left[b_{\mu}(t), b^\dagger_{\mu'}(t')\right] = \delta_{\mu, \mu'} \delta(t-t')$, \blue{where again $\mu = \rm L,R$ correspond to a left/right channel, respectively.}

The noise operators are discretized to be written as matrix product operators (MPOs) acting at each time bin, defined from
\begin{equation}
    \Delta B_{\rm R/L} ^{(\dagger)}  = \int_{t_k}^{t_{k+1}} dt' b_{\rm R/L}^{(\dagger)}(t').
    \label{eq:noise_op}
\end{equation}
These operators create/annihilate a photon in a time bin, following the commutation relation $\left[ \Delta B_{\mu}(t_k), \Delta B_{\mu'}^\dagger(t_{k'}) \right] = \Delta t \delta_{k,k'} \delta_{\mu,\mu'}$. 
They now form a discretized basis,
\begin{equation}
    \ket{i^\mu_k} = \frac{(\Delta B_\mu^\dagger (t_k))^{i^\mu_k}}{\sqrt{i^\mu_k ! (\Delta t)^{i^\mu_k}}} \blue{\ket{0}}.
\end{equation}

Using these operator definitions, we can write Eq.~\eqref{eq:phi_0} in the discretized time basis of length $T$ as 
\begin{widetext}
\begin{equation}
\begin{split}
 \ket{\phi_0^{(1)}} &= \sum_{k=1}^m f_k \Delta B_k^\dagger /\sqrt{\Delta t} \ket{0,...,0} = \sum_{k=1}^m \Delta B_k^\dagger / \sqrt{\Delta }t\ket{0,...,0}_{1,...,m} \otimes \ket{0,...,0}_{m+1,...,T} \\
    &=
   \sum_{\{i_1...i
   _k...i_m\} } A_{a_1}^{(i_1)}A_{a_1,a_2}^{(i_2)}...A_{a_{k-1},a_k}^{(i_k)}...A_{a_{m-2},a_{m-1}}^{(i_{m-1})}A_{a_{m-1}}^{(i_{m})}\ket{0,...,0}_{1,...,m} \otimes \ket{0,...,0}_{m+1,...,T},
\end{split}
\end{equation}
\end{widetext}
where $f_k$ is a unitless discretized version of $f(t)$ contributing to the shape of the pulse at each time bin; $i_k=\{0,1\}$ corresponds to the physical dimension index, which in this case is 0 or 1 photons per time bin, respectively; and $a_k$ corresponds to the additional bond dimensions created by the MPS decomposition. 

In this example with $N=1$, we can write each term explicitly,
as 
\begin{itemize}
    \item if $k=1$:
\begin{equation}
    A_{a_1}^{(0)}=
    \begin{pmatrix}
        1 & 0
    \end{pmatrix}
    , \ A_{a_1}^{(1)}=
    \begin{pmatrix}
        0 & f_1
    \end{pmatrix},
\end{equation}
\item if $1 < k < m$:
\begin{equation}
A_{a_{k-1},a_k}^{(0)}=
    \begin{pmatrix}
         1 & 0 \\
         0 & 1
    \end{pmatrix}, \ A_{a_{k-1},a_k}^{(1)}=
    \begin{pmatrix}
        0 &  f_k \\
        0 & 0
    \end{pmatrix},
\end{equation}
\item if $k=m$:
\begin{equation}
A_{a_{m-1}}^{(0)}=
    \begin{pmatrix}
        0 \\
        1
    \end{pmatrix}, \ A_{a_{m-1}}^{(1)}=
    \begin{pmatrix}
        f_m \\
        0
    \end{pmatrix},
\end{equation}
\end{itemize}

As shown in App.~\ref{app:fock_state_deriv},
this can be generalized for higher photon numbers ($N$) \blue{and an initial state for $N$ photons, $\ket{\phi_0^{(N)}}$, can be calculated using},  
\begin{itemize}
\item If $k=1$:
\begin{align}
    A_{a_1}^{(l)}[i] =& \delta_{l,i}\frac{(f_1)^l}{\sqrt{l!}};
    \label{fock_1}
\end{align}
\item if $1 < k < m$:
\begin{align}
    A_{a_{k-1},a_k}^{(l)}[i,j] =& \delta_{i+l,j}\frac{(f_k)^l}{\sqrt{l!}};
    \label{fock_k}
\end{align}
\item if $k=m$:
\begin{align}
    A_{a_{m-1}}^{(l)}[j] =& \delta_{l,N-j}\frac{(f_m)^l}{\sqrt{l!}};
    \label{fock_m}
\end{align}
\end{itemize}
where $i,j$ are the row/column matrix indices, respectively. Here, we have renamed the physical dimension index $i_k \equiv l$ for simplicity of notation, and $l=\{0,...,N\}$ with $N$ is the number of photons considered. Note also that the normalized total initial state $\ket{\phi_0^{(N)}}$ has a global factor $\sqrt{N!/(\Delta t)^N}$; while these matrices appear different in expression from others seen in the literature \cite{PhysRevA.103.033704,sofia2025}, this is only due to the Gauge freedom of the MPS and they can be shown to be equivalent when subjected to the proper Gauge transformation. Thus, although the matrix forms are different, the computed observables are the same.  

With this time-discretized basis, we can also write the Hamiltonian as an MPO, and define the time evolution operator for each time step, 
\begin{equation}
    U(t_{k+1},t_k) =  \exp{ \left( -i \int_{t_k}^{t_{k+1}} dt' H(t')\right)}.
    \label{eq:t_evol}
\end{equation}

Applying this operator to our initial MPS, we can evolve the entire system through time and extract information from both the TLS and the waveguide photons.

For example, we can calculate the first and second order correlation functions using,
\begin{equation}
    G^{(1)}_{\mu \mu'}(t,t+\tau)= \braket{\Delta B^{\dagger}_{\mu}(t) \Delta B_{\mu'}(t+\tau)}/\Delta t^2
\end{equation}
and
\begin{equation}
\begin{split}
    &G^{(2)}_{\mu \mu'}(t,t+\tau)= \\
    &\braket{\Delta B^{\dagger}_{\mu}(t) \Delta B^{\dagger}_{\mu'}(t+\tau)\Delta B_{\mu'}(t+\tau) \Delta B_{\mu}(t)}/\Delta t^4.
\end{split}
\end{equation}
Note that the correlation functions must be normalized since the basis is not normalized.

\section{Results
for temporal observables with $N$-photon input pulses}
\label{sec:results}

In this section, we start by comparing correlation results from both methods (scattering theory and MPS), using the first-order correlation functions in Sec.~\ref{subsec:g1} and the second-order correlations and population dynamics in Sec.~\ref{subsec:g2}.
We compare results using both 1-photon and 2-photon Fock state pulses. Then, in Sec.~\ref{subsec:pulse_length}, we investigate the role of pulse length and show how it affects the nonlinear effects and populations. Finally, we show the power of MPS in Sec.~\ref{subsec:higher_photons} when using a pulse containing more photons
than two,
and demonstrate results
with up to 8 photons in the Fock-state pulse, 
{yielding a significant nonlinear interaction with 
{\it fully quantized multi-photon pulses}}.

\begin{figure*}[ht]
    \centering
    \includegraphics[width=0.9\textwidth]{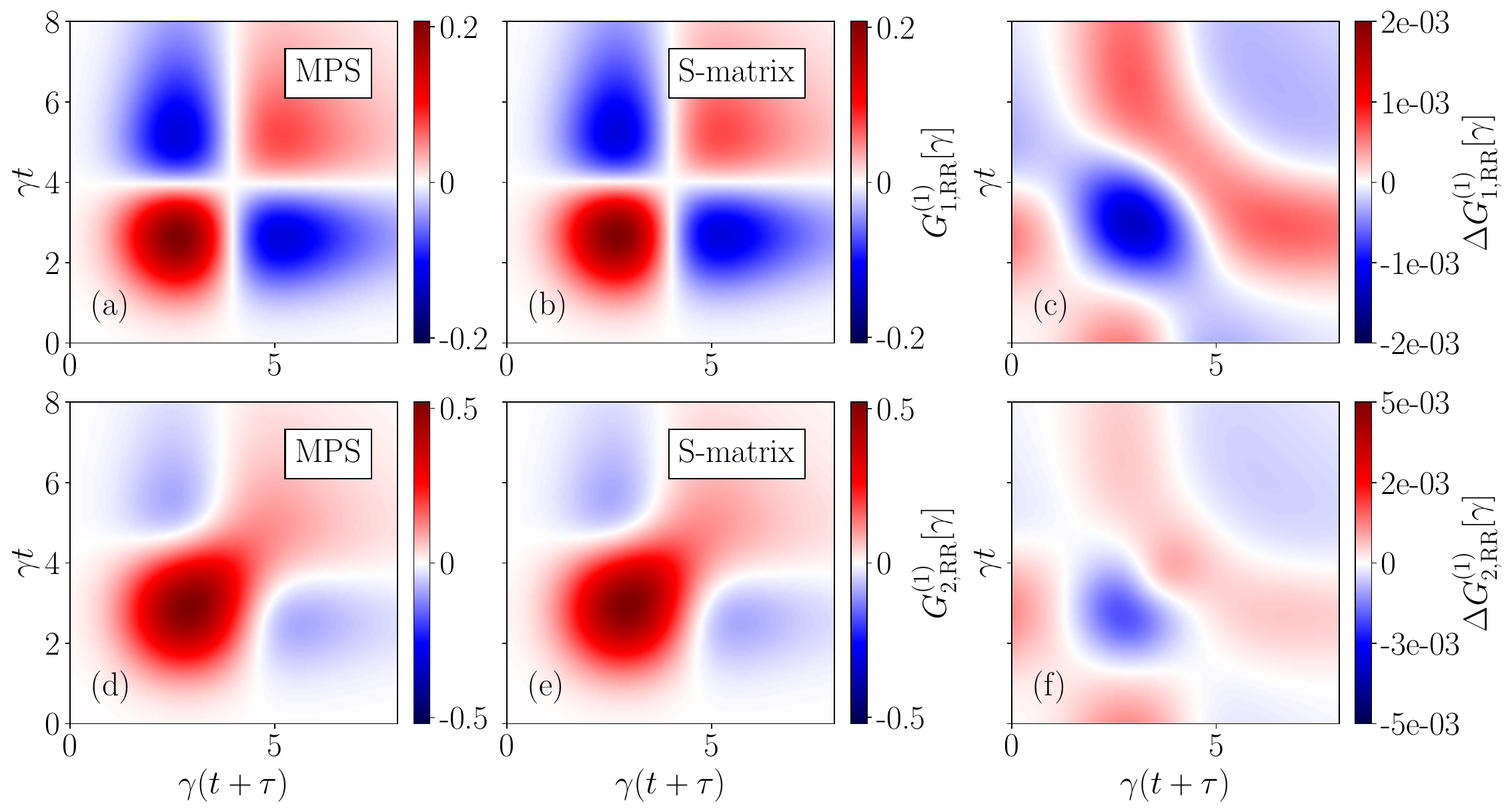}
    \caption{\blue{Two-time first-order quantum correlation functions for the transmitted fields of a TLS symmetrically coupled to a waveguide and driven by a quantum pulse with a Gaussian envelope. Figures (a,b) show the right output of the first-order correlation function with a single-photon pulse, $G^{(1)}_{1,\rm RR}$, calculated using MPS in (a), and the scattering matrix theory in (b). The numerical difference between both approaches, $\Delta G^{(1)}_{\rm 1,RR} = G^{(1)}_{\rm 1,RR} ({\rm MPS}) - G^{(1)}_{\rm 1,RR} ({\rm S{-}matrix})$ is shown in (c). Figures (d,e) show the same first-order correlation function, but in the case of having a pulse containing 2 photons, $G^{(1)}_{2,\rm RR}$, where again both methods are compared, with MPS results in (d) and S-matrix results in (e). Once again, the numerical difference between both approaches, in this case, $\Delta G^{(1)}_{\rm 2,RR}$ is shown in (f).
    MPS calculations were performed with $\Delta t =0.05$ and ${\rm bond_{max}}=8$, and for S-matrix calculations a $\Delta t =0.05$ was considered as well. In both cases, the numerical differences are seen to be very small, and well below 1\% for all two-time points shown}.
    }
    \label{fig:G1}
\end{figure*}

\subsection{First-order two-time correlations}
\label{subsec:g1}

First, we compare the results of 
first-order two-time correlation functions in Fig.~\ref{fig:G1}, where we show results for a quantum pulse with a Gaussian-shaped envelope, which in the time domain is defined from
\begin{equation}
    f(t) = \frac{1}{\pi^{1/4}\sqrt{\sigma_t}} 
    \exp\left \{-\frac{(t-t_c)^2}{2\sigma_t^2}\right \},
\end{equation}
and in the frequency domain,
\begin{equation}
    f(\omega) = \frac{\sqrt{\sigma_t}}{\pi^{1/4}}
    \exp\left \{-\frac{\omega^2\sigma_t^2}{2}\right \}\exp\left \{- i\omega t_c \right \},
    \label{eq:gaussian_w}
\end{equation}
with $t_c$ the center time of the pulse, and $\sigma_t$ the standard deviation in the time domain.
Since we are working 
in the rotating frame at the center frequency of the pulse, then 
$\omega=0$ corresponds to the center of the pulse.

For our chosen example, we center the Gaussian at $\gamma t_c=3$, with a standard deviation of $\gamma \sigma_t=1$, containing one photon [Fig.~\ref{fig:G1}(a,b)] and two photons [Fig.~\ref{fig:G1}(d,e)].
These input pulses interact with a TLS symmetrically coupled to the waveguide. \blue{We compare the results obtained using MPS (labeled with MPS on the figures) and using scattering matrix theory (labeled with S-matrix on the figures) with one and two-photon pulses in Fig.~\ref{fig:G1}(c,f), respectively. In this figure, we show the 
first-order correlation function between right-moving photons at two times, $G^{(1)}_{\rm RR}(t,t+\tau)$. In both cases, we observe quantitatively good agreement between the two methods. Although both methods are in principle exact, small errors appear due to the numerical nature of our implementations, but they remain below one percent for the entire two-time correlation calculations.} 


In Fig.~\ref{fig:G1}(a,b), we observe the center of the Gaussian at $\gamma t_c=3$, where the correlation values get to their maximum. Interestingly, we observe no correlation at $\gamma t \approx 4$ and at $\gamma (t+\tau) \approx 4$, and then we observe correlations, with positive values again when both $t$ and $t+\tau$ are greater than 4, and destructive correlations when only one of them is larger. 

However, when we introduce the 2-photon pulse, in Fig.~\ref{fig:G1}(d,e), we now observe the nonlinear behavior of the two photons interacting, which gives rise to a different first-order correlation. In this case, fewer values of time show no correlation, and constructive interference effects are more present than in the previous case.

\subsection{Second-order two-time correlations and populations}
\label{subsec:g2}

\begin{figure*}[ht]
    \centering
    \includegraphics[width=0.80\textwidth]{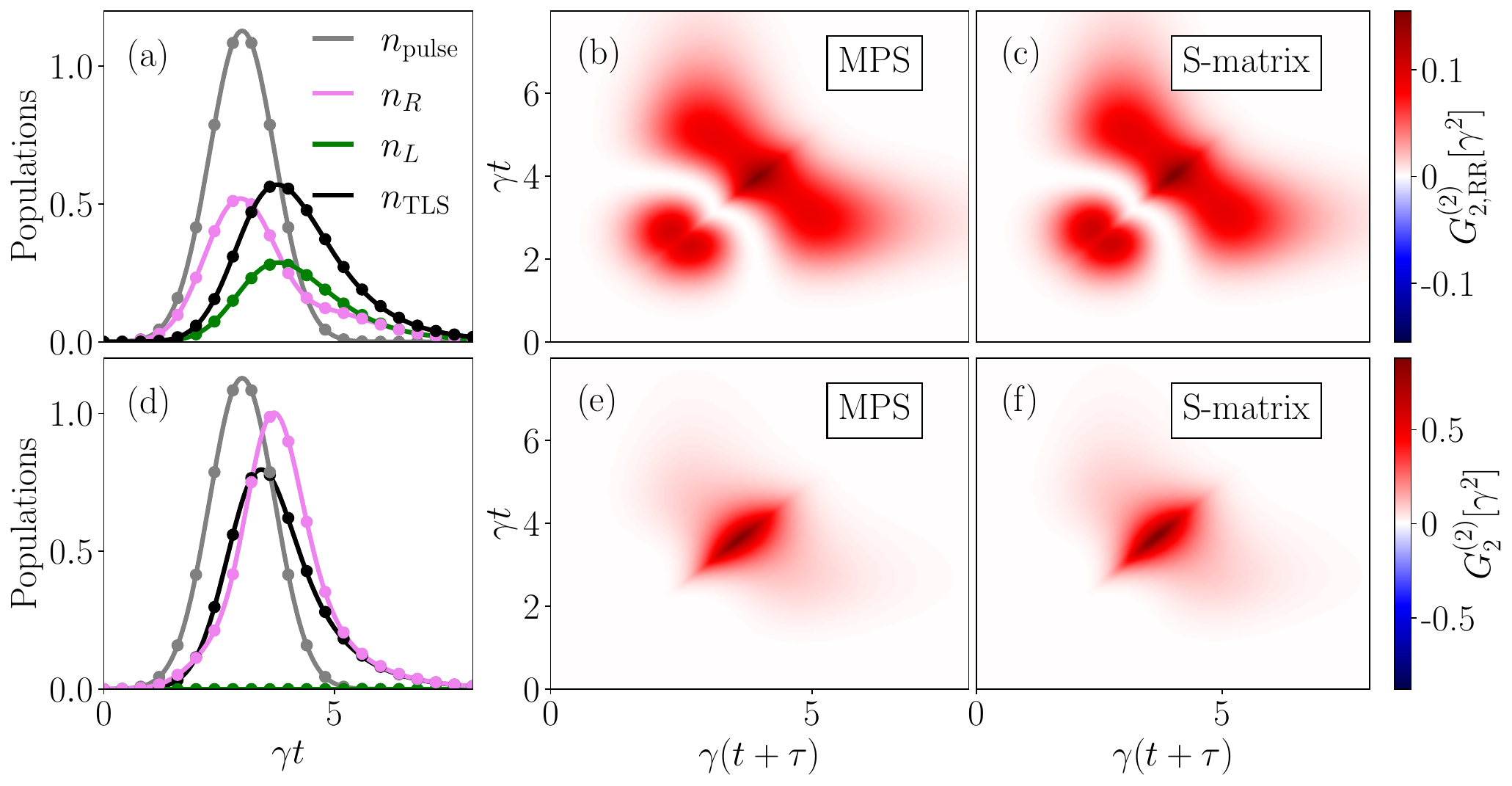}
    \caption{Population dynamics and second-order two-times correlation functions of a 2-photon Gaussian pulse with $\gamma \sigma_t=1$ 
    and centered at $\gamma t_c=3$ 
    interacting with a TLS symmetrically coupled to a waveguide in (a,b,c) and right-chirally coupled to a waveguide in (c,d,e). (a,d) Population dynamics, including the TLS population $n_{\rm TLS}$, input pulse flux $n_{\rm pulse}$, photon flux transmitted to the right $n_{\rm R}$ and reflected to the left $n_{\rm L}$. Solid curves correspond to MPS calculations, and circles correspond to results using S-matrix theory. (b,c) Right channel second-order correlation function, $G^{(2)}_{2,\rm RR} (t,t+\tau)$, comparing MPS results in (b) and S-matrix ones in (c). (e,f) Second-order correlation function in the chiral coupling solution, $G^{(2)}_{2}(t,t+\tau)$, again comparing MPS results in (e) and S-matrix ones in (f). \blue{MPS calculations were performed with $\Delta t =0.05$ and ${\rm bond_{max}}=8$, and for S-matrix calculations a $\Delta t =0.05$ was considered as well}.
    }
    \label{fig:G2pops}
\end{figure*}
\begin{figure*}
    \centering
    \includegraphics[width=0.96\textwidth]{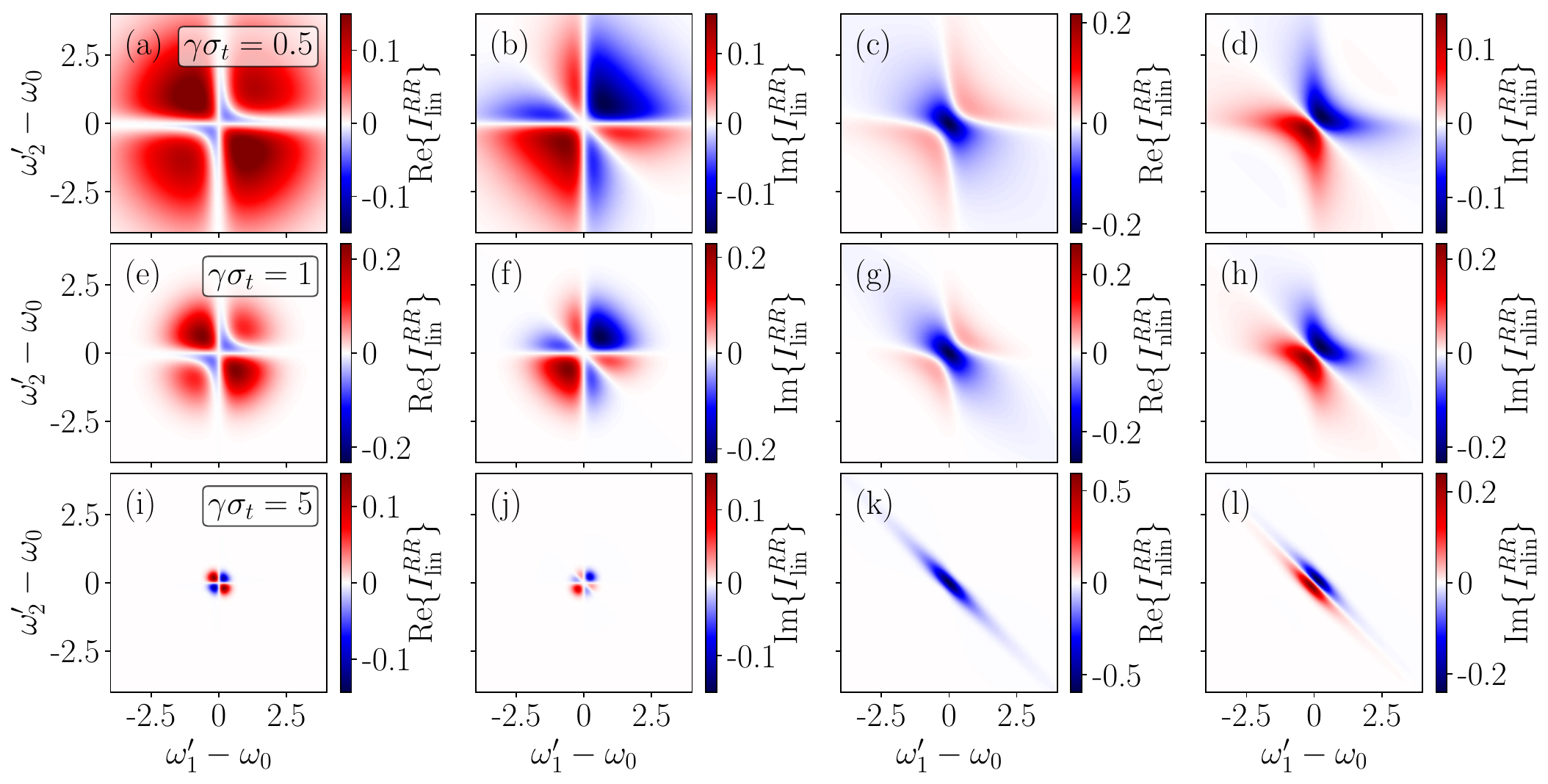}
    \caption{Study of $I^{RR}_{\rm lin}$ and $I^{\rm sym}_{\rm nlin}$ components on the transmission (Eqs.~\eqref{Ilin} and~\eqref{Inlin}, respectively) for a 2-photon Gaussian pulse with $\sigma_t=0.5$ in (a,b,c,d), $\sigma_t=1$ in (e,f,g,h), and $\sigma_t=5$ in (i,j,k,l), interacting with TLS symmetrically coupled to a waveguide. The real part of the linear component $I^{RR}_{\rm lin}$ is shown in (a,e,i), and the imaginary part in (b,f,j). The real parts of the nonlinear component $I^{\rm sym}_{\rm nlin}$ is shown in (c,g,k) and the imaginary parts in (d,h,l).}
    \label{fig:ilin_inlin_v2}
\end{figure*}

Figure~\ref{fig:G2pops} presents the population dynamics of a 2-photon Gaussian pulse with a standard deviation $\gamma \sigma_t=1$, and centered at $\gamma t_c=3$ interacting with a TLS coupled to a waveguide. 

In Fig.~\ref{fig:G2pops}(a,b,c), we show results for a symmetrically coupled TLS, and then in Fig.~\ref{fig:G2pops} (d,e,f), the same observables are shown for a right-chiral coupled TLS. Population dynamics are calculated using the two methods described before, including the input pulse photon flux ($n_{\rm pulse}$), the photon flux transmitted to the right ($n_R$), the photon flux reflected to the left ($n_L$) and the TLS population ($n_{\rm TLS}$). As expected, the symmetrical solution shows both right and left output fluxes, while in the chiral case, there is only transmission, and the left flux (shown in green in the Figure) remains zero. In addition, the TLS gets more excited when chirally coupled, with a larger maximum value of the TLS population. 

As before, {\it \blue{quantitatively good} agreement is shown between scattering theory and MPS}, both in the symmetrical and chiral solutions; solid lines correspond to MPS calculations, while circles represent the scattering matrix solutions. This demonstrates that the TLS population can indeed be calculated using Eq.~\eqref{eq:tls_pop_f}, even in the nonlinear regime.

The two-time second-order correlation functions are calculated again using both techniques and, not surprisingly, show agreement once more. \blue{Note here that similar levels of agreement to the ones shown in Fig.~\ref{fig:G1} are obtained, but difference graphs are not explicitly shown.}
In the symmetrical solution, we calculate the correlation between the right output channels [$G^{(2)}_{\rm 2,RR}(t,t+\tau)$] and observe the characteristic bird-like shape. In the chiral solution, there is only one channel, and the chiral $G^{(2)}_2(t,t+\tau)$ is calculated, showing the maximum values of correlation when $\tau=0$, i.e., when both times coincide.

\subsection{The role of pulse length and quantum nonlinearities}
\label{subsec:pulse_length}

\begin{figure*}[ht]
    \centering
    \includegraphics[width=0.9\textwidth]{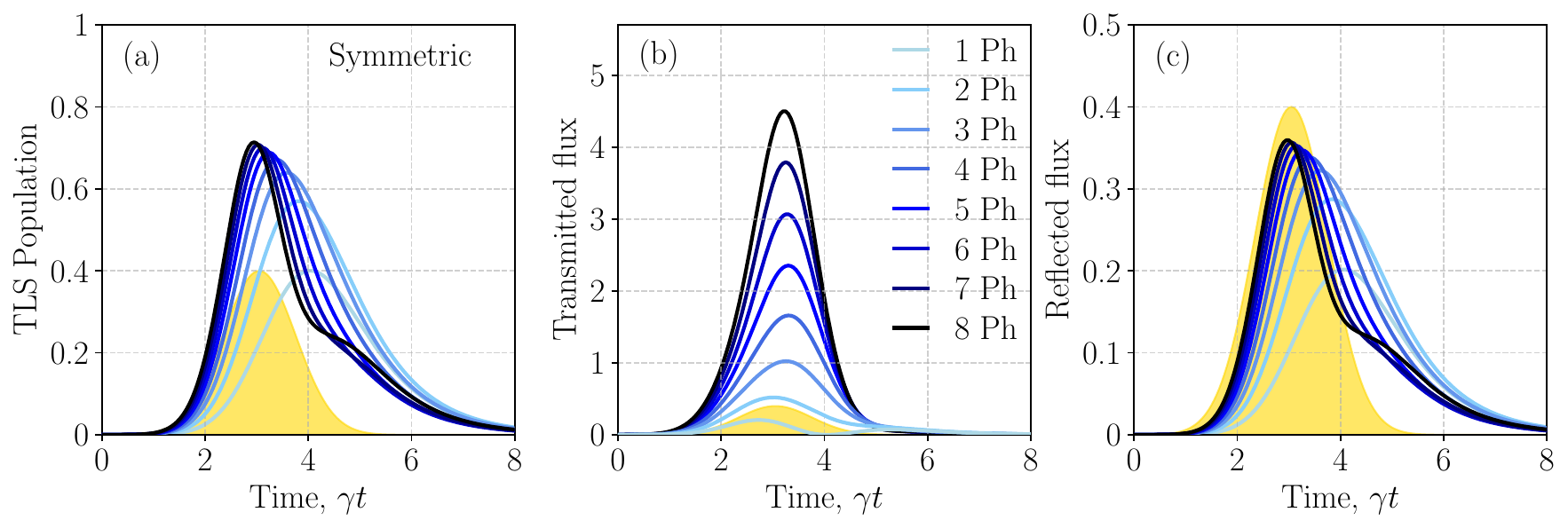}
    \includegraphics[width=0.65\textwidth]{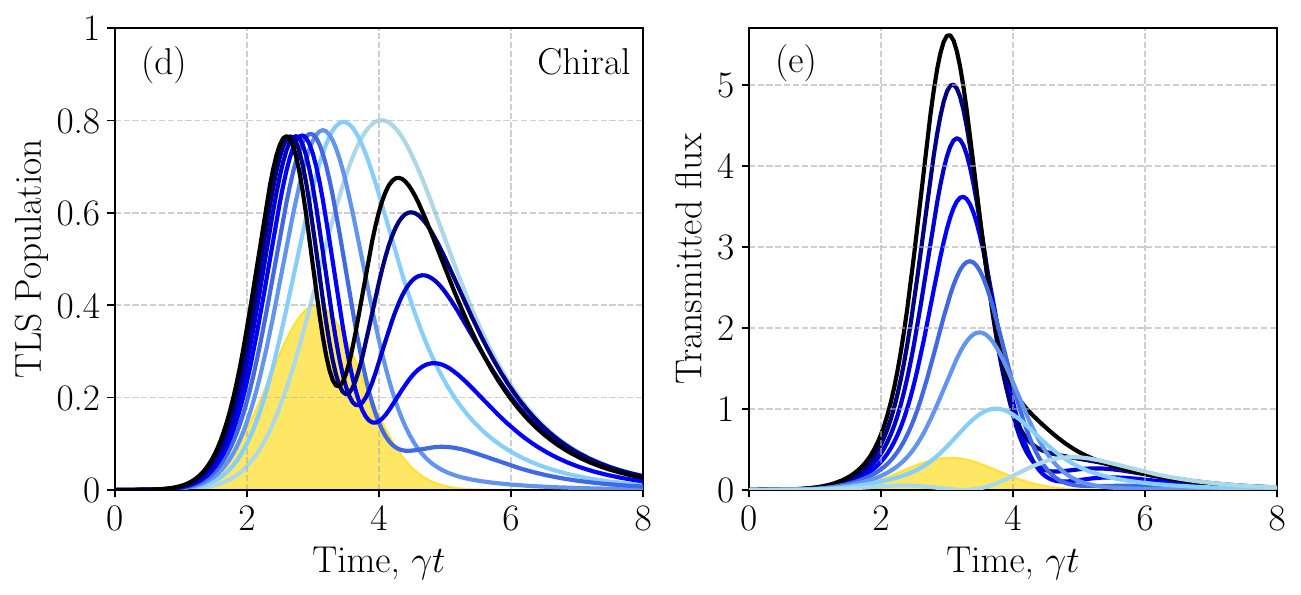}
    \caption{Population dynamics of an $n$-photon Gaussian Fock-state pulse with $\gamma \sigma_t=1$ and centered at $\gamma t_c=3$, interacting with a TLS symmetrically coupled (a,b,c) and chirally coupled (d,e) to a waveguide. (a,d) TLS population dynamics. (b,e) Transmitted photon flux, and (c) reflected photon flux. The input pulse is shown in yellow for reference (renormalized to a maximum of 0.4).
    }
    \label{fig:high_ph_num_ch}
\end{figure*}

One advantage of using scattering theory is the fact that one has more control over the different parts of the analytical derivation, with the ability to directly see and study intermediate steps in the frequency domain, which can give additional insights into the light-matter interaction in different cases.

For example, fundamental nonlinear dynamics appear when working with pulses containing two-photons. This can be directly understood in 
Eq.~\eqref{eq:2ph_proj}, where the new terms $I_{\rm lin}$ and $I_{\rm nlin}$ contribute to the linear and nonlinear interactions, respectively. These terms are defined in Eqs.~\eqref{Ilin} and~\eqref{Inlin}, and studied in Fig.~\ref{fig:ilin_inlin_v2} for different pulse lengths. 

To investigate the role of pulse
duration, we investigate three different pulse lengths: $\gamma \sigma_t=0.5$ [Fig.~\ref{fig:ilin_inlin_v2}(a-d)], $\gamma \sigma_t=1$ [Fig.~\ref{fig:ilin_inlin_v2}(e-h)] and $\gamma \sigma_t=5$ [Fig.~\ref{fig:ilin_inlin_v2}(i-l)]. In each case, the real and imaginary contributions
to $I_{\rm lin},I_{\rm nlin}$
are studied. For these simulations, we choose the pulse centered at $\gamma t_c=0$, since in the frequency domain changing the pulse center only adds interferences [see Eq.~\eqref{eq:gaussian_w}].

We start by considering the linear regime, where we observe that the longer the pulse length in the time domain, the narrower the interaction will be in the frequency domain. This can be understood by inspecting Eq.~\eqref{Ilin}, where we see that if we focus on the transmission, the linear term is
\begin{equation}
    I^{RR}_{\rm lin}(\omega_1,\omega_2) = f(\omega_1,\omega_2) t_{\rm sym} (\omega_1) t_{\rm sym} (\omega_2),
\end{equation}
which is simply a multiplication of the transmission coefficients and the pulse envelope (causing convolutions in the time domain).
We have chosen to use an example for the symmetric emitter, but the general arguments also apply to the chiral case.
Consequently, 
Fig.~\ref{fig:ilin_inlin_v2}, panels (a,b) have the widest interferences, since these shorter pulses in time ($\gamma\sigma_t=0.5$) have the largest frequency bandwidth. 
It is also important to note here that we are working in the interaction picture,
at the rotating frame $\omega_0$; to make this clear, in Fig.~\ref{fig:ilin_inlin_v2}, we define $\omega^\prime_{1,2}-\omega_0\equiv\omega_{1,2}$. 

If we now focus on the real component shown in Fig.~\ref{fig:ilin_inlin_v2}(a), positive interferences appear in all quadrants, and negative ones are only created in quadrants with frequencies having the same sign, which indicates linear transmission coefficients with the same or opposite signs, respectively. A similar behavior is shown 
in Fig.~\ref{fig:ilin_inlin_v2}(e), but now with a narrower range of frequencies. However, when the pulse is long enough, such as 
in Fig.~\ref{fig:ilin_inlin_v2}(i), the interaction narrows down, and the positive components do not form on the quadrants with the same frequency sign.  

Additionally, this term goes to zero when one of the frequencies is on-resonance with the natural frequency ($\omega_0$), which means that at least one of the transmission coefficients is zero, and the linear component is purely real when the frequencies have opposite signs, indicating that the transmission coefficients in this case do not have complex components, or they cancel out. These terms could help us extract information known from linear optics, such as a pulse phase change produced during the interaction; however, to have a full picture of the interaction, we need to first also consider the nonlinear term, which in this case is $I^{\rm sym}_{\rm nlin}$.  

If we now study the nonlinear term $I^{\rm sym}_{\rm nlin}$ from Eq.~\eqref{Inlin},
then (again using the symmetric emitter case)
\begin{equation}
\begin{split}
&I^{\rm sym}_{\rm nlin}(\omega_1,\omega_2) =
\frac{1}{2}\int d \Delta' f(\omega'-\Delta',\omega'+\Delta') \times \\
&\frac{4}{\pi \gamma} \frac{r_{\rm sym}(\omega' - \Delta')r_{\rm sym}(\omega' + \Delta') r_{\rm sym}(\omega_1)r_{\rm sym}(\omega_2)}{r_{\rm sym}\left(\frac{\omega_1+\omega_2}{2}\right)},
\label{Irr}
\end{split}
\end{equation}
where, as defined before, $\omega' = \frac{\nu_1 + \nu_2}{2}$ and $ \Delta'=\frac{\nu_2 - \nu_1}{2}$. 
We can see how the interaction is not simply a multiplication in frequency anymore, with a more complicated relationship between the different terms of the nonlinear components. This is because the additional constraint in the nonlinear component is the energy conservation, in contrast to the linear regime, where no new frequencies are generated. Hence, we can now observe the generation of new frequencies via the nonlinear interaction.

We note that for the nonlinear contribution, the real terms behave differently from the linear parts, with values being different from zero at the resonant frequencies, and canceling at different frequencies that depend on the pulse length. We can no longer separate the pulse behavior from the linear transmission and reflection coefficients, since Eq.~\eqref{Irr} is an integral presenting a convolution of the pulse with the rest of the other terms. However, we can follow a similar interference behavior, with positive interferences closer to the second and fourth quadrants, and a negative one in the first and third, when the interaction is stronger [Fig.~\ref{fig:ilin_inlin_v2}, panels (c) and (g)]; and, again, with the longest pulse [Fig.~\ref{fig:ilin_inlin_v2}(k)], the behavior changes, with now only a negative contribution following the diagonal of opposite signed frequencies. The corresponding imaginary contributions also approach a diagonal tendency for longer pulses, suggesting that this is a characteristic of the weaker interaction. Thus, results closer to the first two cases show a strong nonlinear interaction, while the third row of results presented for a longer pulse shows a weak interaction with the quantum emitter.   

It is important to note here that this information cannot be extracted in the time observables such as the first and second order correlation functions [Eqs.~\eqref{G1_2ph} and~\eqref{G2_2ph}] since in all the observables the projection is multiplied by a complex conjugate.   

This behavior helps us estimate the strength of the interaction and understand the influence of considering the pulse wavepacket. Most of the scattering studies have been done disregarding the pulse envelope, i.e. in the limit when we have a Dirac delta function. Here, we observe how the interaction is more complicated when considering bandwidth pulses, where both frequencies interfere with an intrinsic dependence on the pulse shape and duration.

\subsection{Study of TLS interactions with  higher photon number Fock-state pulses,
with $N=1$ to $N=8$}
\label{subsec:higher_photons}

Finally, in this subsection, we show the significant power of MPS by extending our study to pulses containing more photons, which is a significant challenge
and typically not possible with other approaches, such as scattering matrix techniques. 
\blue{
This regime has been outlined using the Heisenberg approach in Ref.~\cite{PhysRevA.90.063832}, but they are limited to emitter dynamics, and results for more than two photons were not calculated.  

Usually, introducing higher numbers of photons is limited to coherent input states in which the Heisenberg-Langevin equations can be reduced to a closed ODE form~\cite{PhysRevA.83.063842}. However, here we show how our approach can handle an input Fock-state pulse containing multiple photons with selected examples up to $N=8$. }

To showcase this, in Fig.~\ref{fig:high_ph_num_ch}, we calculate the population dynamics of the interaction of Gaussian pulses containing up to 8 photons, in a Fock state, with a TLS coupled to a waveguide, symmetrically on the top row  [Fig.~\ref{fig:high_ph_num_ch} (a,b,c)], and chirally on the bottom row [Fig.~\ref{fig:high_ph_num_ch} (d,e)]. In all cases, the pulse envelope is centered at $\gamma t_c = 3$ and has a standard deviation $\gamma\sigma_t=1$.

In the symmetrical solution, we observe how the population increases with the photon number in Fig.~\ref{fig:high_ph_num_ch}(a), as well as its decay speed due to the presence of stimulated emission when dealing with more photons. In panels (b) and (c) of Fig.~\ref{fig:high_ph_num_ch}, the transmitted and reflected fluxes are presented. The transmitted flux shows a high growth as the photon number increases. Interestingly, the reflected flux follows exactly the same dynamics as the TLS population, except for a factor of two, i.e., for a right propagating input pulse, $n_{L} = n_{\rm TLS}/2$.
This can be analytically seen by using Eq.~\eqref{heis3}, and considering that for the left channel, the waveguide is originally in vacuum since the input pulse is traveling towards the right, this leads to,
\begin{equation}
    n_{L} = \braket{a_{\rm out}^{\rm L \ \dagger}(t)a_{\rm out}^{\rm L}(t)}=\gamma/2 \braket{\sigma^+\sigma^-}(t) = \frac{\gamma}{2}  n_{\rm TLS},
\end{equation}
which in units of $\gamma$ transforms to $n_{L} = n_{\rm TLS}/2$.

The chirally coupled TLS presents completely different dynamics [Fig.~\ref{fig:high_ph_num_ch}(d)], with a maximum that not only does not increase with the photon number, but is even slightly decreasing. In addition, the stimulated emission is much more present, giving rise to a peak splitting that is visible with photon numbers as low as 4. As expected, the transmitted flux [Fig.~\ref{fig:high_ph_num_ch} (e)] rapidly increases with the photon number, with a higher rise than in the symmetrical solution since there is no reflection, and in this case, the dynamics become faster, with a narrowing of the flux. 

\blue{
These nonlinear population oscillations during the pulse interaction, appearing for higher numbers of photons, show how the system approaches the strong pumping regime where Rabi-like oscillations become present, but in this case, with quantum input fields that have a vanishing electric field expectation value. As more photons are introduced to the system, the dynamics become closer to the classical limit. Similar strong-field phenomenon has been previously reported in Refs.~\cite{Mukhopadhyay_2024,PhysRevA.88.013855}.
}

\section{Conclusions}
\label{sec:conclusions}


We have presented a detailed theory with simulation results to study the 
quantum light-matter dynamics with Fock-state pulses, containing $N$ photons, interacting with a single TLS coupled to a waveguide. In the regime of one and two-photon pulses, we started by deriving the relevant equations and methods with two powerful and well-established theoretical techniques: (i) scattering matrix theory, and (ii) matrix product states.

We began by showing the derivation of the scattering matrices for the cases of one-photon and two-photon input pulses of an arbitrary temporal shape. We presented derivations of the interaction with a symmetrically coupled TLS as well as a chirally coupled one. In addition, we calculated the 
two-time first-order field correlation functions
and the second-order field correlations (in the case of having a two-photon input pulse). Moreover, we have shown a way to extract the TLS populations, something that is not usually considered with the scattering matrix method.

Then we have introduced MPS, where we have given a general form of a Fock state input pulse containing $N$ photons, and presented field correlation function equations written in terms of MPOs.

In the results section, we have shown a direct comparison of both methods, with \blue{quantitatively good} agreement in all the results presented, which included first- and second-order two-time correlation functions for interactions with one-photon and two-photon Gaussian pulses, both when the TLS is symmetrically and chirally coupled. In addition, we have shown the population dynamics, including the photon fluxes and the emitter population, with again agreement between the two approaches. The general physics of these two different methods is seen as complementary. Moreover, both methods are exact, or more correctly, numerically exact.

Later, we focused on the scattering matrix method again to study the linear and nonlinear components of the interaction of a two-photon Gaussian pulse in the frequency domain, and discussed how this impacts temporal dynamics. For this, we have calculated the real and imaginary components of both terms of the projected wavefunction of two photons, $I^{RR}_{\rm lin}$ and $I^{\rm sym}_{\rm nlin}$, respectively, and studied how they behave for three different pulse lengths. 

Here, we showed that when considering the bandwidth of the pulse, the nonlinear interaction becomes a convolution, where one can no longer separate the linear transmission and reflection components from the pulse envelope to extract direct information (such as a phase shift) as is usually done in the linear regime. 
This is captured in the real and imaginary components of $I^{\rm sym}_{\rm nlin}$, where we observed complex interference effects. These effects show up in a different way when calculating temporal correlations, where the projection of the wavefunction loses its complex nature [Eq.~\eqref{G2_2ph}], and the known bird-like second-order correlation feature is recovered on a two-time map. 

In addition, we have observed how, when the pulse is longer in the time domain, the nonlinear term behaves qualitatively differently, showing a weaker interaction with the TLS. 

Next, we have shown one of the significant advantages of using MPS, which is the ability to study interactions with higher numbers of (quantized) photons,  with efficient numerical calculations that are usually inaccessible due to the increasing Hilbert space. We have demonstrated this by studying the interaction with Gaussian Fock-state pulses going from 1 to 8 photons. We have studied both symmetrical and chiral interactions, presenting quantum dynamical results, including TLS populations and transmitted and reflected fluxes. We also have observed how, in the symmetrical case, the reflected fluxes will follow the same dynamics as the TLS population, apart from a $\gamma/2$ factor. In the chiral dynamics, we have observed very interesting TLS population dynamics, showing signs of stimulated emission and reabsorption with a peak splitting as the photon number increases, with clear evidence in cases where pulses contain four photons or more.

In terms of applications, broadly, this work has shown a
powerful time-dependent study with both techniques, with full derivations; we gave an analysis and discussed the advantages of using each method, shown through specific results, which gives a deeper understanding of the subject. 
\blue{For instance, we have shown the advantage of having direct access to the frequency wavefunction with the S-matrix approach. This becomes important when working with highly correlated two-photon input states. Such states are more complicated to implement using MPS, and the maximum bond dimension required to model the state, increases with the growing entanglement between bins.}

Although we have focused on Gaussian pulses, both techniques can be used with other pulse envelopes, and other cases, such as imperfect coupling, can be studied as well. Additionally, MPS is not limited to 8 photons, and higher numbers can 
easily be used for other studies
as well. \blue{A further MPS advantage is the ability to include losses, such as pure dephasing and off-chip decay, when converted to a density matrix approach~\cite{bundgaardnielsen2026}, whereas a clear implementation has not been developed using S-matrix techniques.}

All the MPS results shown in this work were calculated using the
recently developed open-source Python package QwaveMPS~\cite{2602.15826}.

\acknowledgements
This work was supported by the Natural Sciences and Engineering Research Council of Canada (NSERC),
the National Research Council of Canada (NRC),
the Canadian Foundation for Innovation (CFI), and Queen's University, Canada.
We thank Jacob Ewaniuk and
Nir Rotenberg for useful discussions.

\appendix

\section{Derivation of Fock State Matrices}

\label{app:fock_state_deriv}

In this appendix, we will briefly show how to derive the matrix elements for the Fock state used in the main text [Eqs.~\eqref{fock_1},~\eqref{fock_k} and~\eqref{fock_m}],
\begin{align}
    A_{a_1}^{(l)}[i] =& \delta_{l,i}\frac{(f_1)^l}{\sqrt{l!}},\\
    A_{a_{k-1},a_k}^{(l)}[i,j] =& \delta_{i+l,j}\frac{(f_k)^l}{\sqrt{l!}},\\
    A_{a_{m-1}}^{(l)}[j] =& \delta_{l,N-j}\frac{(f_m)^l}{\sqrt{l!}},
\end{align}
with $i,j$ the row/column matrix indices, $l$ the physical dimension index, where $l=\{0,...,N\}$, and $N$ is the number of photons considered. The normalized total initial state $\ket{\phi_0^{(N)}}$ has a global factor $\sqrt{N!/(\Delta t)^N}$. 

This derivation is based on the ideas discussed in Ref.~\cite{bacon2008nfa}, approaching the creation of the matrices from the perspective of formal language theory over the alphabet of the number of photons that might be placed in each individual bin. The main idea of the process is, since we are working in the Fock state basis, to use the bond indices to track the number of photons that have been placed in the MPS chain up to a given point, analogous to the memory of a finite automaton. 
In this sense of formal language theory, we think of the MPS as encoding the complex coefficient associated with a given sequence of photon numbers in bins.

We begin by considering a Fock state described as a product state of the different temporal envelopes of the photons. We then have the state
\begin{align}
    \ket{\phi_0^{(N)}} =& \frac{1}{\sqrt{N!}} \prod_{i=1}^N\left(\int dt_i f^{(i)}(t_i)a^\dag(t_i)\right)\ket{0} \nonumber\\
    \overset{\rm discretize}{\rightarrow}=& \frac{1}{\sqrt{N!}} \prod_{i=1}^N \left(\sum_{j=1}^m f^{(i)}(t_j)\Delta B^\dag(t_j)/\sqrt{\Delta t} \right)\ket{0}. 
    \label{eq:discretized_fock}
\end{align}

If we assume identical/indistinguishable photons, we have that $f^{(i)} = f, \forall i$. We can also shorten the notation by letting $f_j \equiv f(t_j)$ and $\Delta B^\dag_j \equiv \Delta B^\dag(t_j)$. Doing so, we have:
\begin{align}
    \ket{\phi_0^{(N)}} =& \frac{1}{\sqrt{N!}} \prod_{i=1}^N \left(\sum_{j=1}^mf_j\Delta B^\dag_j/\sqrt{\Delta t}\right)\ket{0} \nonumber\\
    =& \frac{1}{\sqrt{N!}} \left(\sum_{j=1}^m f_j\Delta B^\dag_j/\sqrt{\Delta t}\right)^N\ket{0},
\end{align}
where $\ket{0}$ represents the vacuum state of the waveguide.
It is then useful to apply the multinomial expansion, which states that
\begin{align}
    \left(\sum_{j=1}^m x_j\right)^N = \sum_{\{n_k\}} \frac{N!}{\prod_{k=1}^m n_k!} \prod_{k=1}^m x_j^{n_k},
\end{align}
where the sum over $\{n_k\}$ is the sum over all possible partitions of $N$ over the $m$ different bins and as such we require that $\sum_k n_k = N$.

Using this result it follows that
\begin{align}
    \ket{\phi_0^{(N)}} =& \frac{1}{\sqrt{N!(\Delta t)^N} } \sum_{\{n_k\}} \frac{N!}{\prod_{k=1}^m n_k!} \prod_{k=1}^m (f_k \Delta B^\dag_k)^{n_k} \ket{0} \nonumber\\
    =& \frac{1}{\sqrt{N!}} \sum_{\{n_k\}} \frac{N!\sqrt{\prod_{k=1}^m n_k!}}{\prod_{k=1}^m n_k!} \bigotimes_{k=1}^m (f_k)^{n_k} \ket{(n_k)_k} \nonumber\\
    =& \sqrt{N!} \sum_{\{n_k\}} \sqrt{\frac{1}{\prod_{k=1}^m n_k!}} \bigotimes_{k=1}^m (f_k)^{n_k} \ket{(n_k)_k},
\end{align}
where $\ket{(n_k)_k}$ denotes $n_k$ photons in the $k^{\rm th}$ site of the discretized waveguide, in the second step we operate on the vacuum state at each site with $(\Delta B_k^\dag)^{n_k}$; in the final step we factor out the factor of $\sqrt{N!}$ to simplify the final result.

This equation describes the amplitude associated with each partition of the $N$ photons among the $m$ bins of the length of the pulse. Each has a relative amplitude of $\sqrt{\frac{1}{\prod_{k=1}^m n_k!}} \prod_{k=1}^m f_k^{n_k}$ for each partition.

Motivated by formal language theory, we find it convenient to consider the change in the amplitude as we append one more site to the chain (in the terminology of formal language theory, one more letter to the word). To do so, we consider the partial chain of length $l$, which has an amplitude of
\begin{align}
    C_l(n_1,n_2,\dots,n_l) = \sqrt{N!}\sqrt{\frac{1}{\prod_{k=1}^l n_k!}} \prod_{k=1}^l (f_k)^{n_k}.
\end{align}

We can then see that the relative change from appending one more site/letter to the chain of length $l-1$ is then given by
\begin{align}
    R_{l} \equiv \frac{C_{l}(n_1,n_2,\dots,n_{l-1},n_{l})}{C_{l-1}(n_1,n_2,\dots,n_{l-1})} = \frac{f_{l}^{n_l}}{\sqrt{n_{l}!}}.
\end{align}
This $R_{l}$ can be thought of as a transition amplitude associated with appending a site with photon number $n_{l}$, and can be used to construct the matrices either directly or indirectly via the construction of a complex weighted finite state automata \cite{bacon2008nfa}.


\bibliography{references}

\begin{thebibliography}{70}%
\makeatletter
\providecommand \@ifxundefined [1]{%
 \@ifx{#1\undefined}
}%
\providecommand \@ifnum [1]{%
 \ifnum #1\expandafter \@firstoftwo
 \else \expandafter \@secondoftwo
 \fi
}%
\providecommand \@ifx [1]{%
 \ifx #1\expandafter \@firstoftwo
 \else \expandafter \@secondoftwo
 \fi
}%
\providecommand \natexlab [1]{#1}%
\providecommand \enquote  [1]{``#1''}%
\providecommand \bibnamefont  [1]{#1}%
\providecommand \bibfnamefont [1]{#1}%
\providecommand \citenamefont [1]{#1}%
\providecommand \href@noop [0]{\@secondoftwo}%
\providecommand \href [0]{\begingroup \@sanitize@url \@href}%
\providecommand \@href[1]{\@@startlink{#1}\@@href}%
\providecommand \@@href[1]{\endgroup#1\@@endlink}%
\providecommand \@sanitize@url [0]{\catcode `\\12\catcode `\$12\catcode `\&12\catcode `\#12\catcode `\^12\catcode `\_12\catcode `\%12\relax}%
\providecommand \@@startlink[1]{}%
\providecommand \@@endlink[0]{}%
\providecommand \url  [0]{\begingroup\@sanitize@url \@url }%
\providecommand \@url [1]{\endgroup\@href {#1}{\urlprefix }}%
\providecommand \urlprefix  [0]{URL }%
\providecommand \Eprint [0]{\href }%
\providecommand \doibase [0]{https://doi.org/}%
\providecommand \selectlanguage [0]{\@gobble}%
\providecommand \bibinfo  [0]{\@secondoftwo}%
\providecommand \bibfield  [0]{\@secondoftwo}%
\providecommand \translation [1]{[#1]}%
\providecommand \BibitemOpen [0]{}%
\providecommand \bibitemStop [0]{}%
\providecommand \bibitemNoStop [0]{.\EOS\space}%
\providecommand \EOS [0]{\spacefactor3000\relax}%
\providecommand \BibitemShut  [1]{\csname bibitem#1\endcsname}%
\let\auto@bib@innerbib\@empty
\bibitem [{\citenamefont {Hughes}(2004)}]{Hughes2004}%
  \BibitemOpen
  \bibfield  {author} {\bibinfo {author} {\bibfnamefont {S.}~\bibnamefont {Hughes}},\ }\bibfield  {title} {\bibinfo {title} {Enhanced single-photon emission from quantum dots in photonic crystal waveguides and nanocavities},\ }\href {https://doi.org/10.1364/ol.29.002659} {\bibfield  {journal} {\bibinfo  {journal} {Optics Letters}\ }\textbf {\bibinfo {volume} {29}},\ \bibinfo {pages} {2659} (\bibinfo {year} {2004})}\BibitemShut {NoStop}%
\bibitem [{\citenamefont {Zheng}\ \emph {et~al.}(2010)\citenamefont {Zheng}, \citenamefont {Gauthier},\ and\ \citenamefont {Baranger}}]{Zheng2010}%
  \BibitemOpen
  \bibfield  {author} {\bibinfo {author} {\bibfnamefont {H.}~\bibnamefont {Zheng}}, \bibinfo {author} {\bibfnamefont {D.~J.}\ \bibnamefont {Gauthier}},\ and\ \bibinfo {author} {\bibfnamefont {H.~U.}\ \bibnamefont {Baranger}},\ }\bibfield  {title} {\bibinfo {title} {Waveguide {QED}: Many-body bound-state effects in coherent and {Fock-state} scattering from a two-level system},\ }\href {https://doi.org/10.1103/physreva.82.063816} {\bibfield  {journal} {\bibinfo  {journal} {Phys. Rev. A}\ }\textbf {\bibinfo {volume} {82}},\ \bibinfo {pages} {063816} (\bibinfo {year} {2010})}\BibitemShut {NoStop}%
\bibitem [{\citenamefont {Longo}\ \emph {et~al.}(2011)\citenamefont {Longo}, \citenamefont {Schmitteckert},\ and\ \citenamefont {Busch}}]{PhysRevA.83.063828}%
  \BibitemOpen
  \bibfield  {author} {\bibinfo {author} {\bibfnamefont {P.}~\bibnamefont {Longo}}, \bibinfo {author} {\bibfnamefont {P.}~\bibnamefont {Schmitteckert}},\ and\ \bibinfo {author} {\bibfnamefont {K.}~\bibnamefont {Busch}},\ }\bibfield  {title} {\bibinfo {title} {Few-photon transport in low-dimensional systems},\ }\href {https://doi.org/10.1103/PhysRevA.83.063828} {\bibfield  {journal} {\bibinfo  {journal} {Phys. Rev. A}\ }\textbf {\bibinfo {volume} {83}},\ \bibinfo {pages} {063828} (\bibinfo {year} {2011})}\BibitemShut {NoStop}%
\bibitem [{\citenamefont {Longo}\ \emph {et~al.}(2010)\citenamefont {Longo}, \citenamefont {Schmitteckert},\ and\ \citenamefont {Busch}}]{PhysRevLett.104.023602}%
  \BibitemOpen
  \bibfield  {author} {\bibinfo {author} {\bibfnamefont {P.}~\bibnamefont {Longo}}, \bibinfo {author} {\bibfnamefont {P.}~\bibnamefont {Schmitteckert}},\ and\ \bibinfo {author} {\bibfnamefont {K.}~\bibnamefont {Busch}},\ }\bibfield  {title} {\bibinfo {title} {Few-photon transport in low-dimensional systems: Interaction-induced radiation trapping},\ }\href {https://doi.org/10.1103/PhysRevLett.104.023602} {\bibfield  {journal} {\bibinfo  {journal} {Phys. Rev. Lett.}\ }\textbf {\bibinfo {volume} {104}},\ \bibinfo {pages} {023602} (\bibinfo {year} {2010})}\BibitemShut {NoStop}%
\bibitem [{\citenamefont {Rom\'an-Roche}\ \emph {et~al.}(2020)\citenamefont {Rom\'an-Roche}, \citenamefont {S\'anchez-Burillo},\ and\ \citenamefont {Zueco}}]{PhysRevA.102.023702}%
  \BibitemOpen
  \bibfield  {author} {\bibinfo {author} {\bibfnamefont {J.}~\bibnamefont {Rom\'an-Roche}}, \bibinfo {author} {\bibfnamefont {E.}~\bibnamefont {S\'anchez-Burillo}},\ and\ \bibinfo {author} {\bibfnamefont {D.}~\bibnamefont {Zueco}},\ }\bibfield  {title} {\bibinfo {title} {Bound states in ultrastrong waveguide {QED}},\ }\href {https://doi.org/10.1103/PhysRevA.102.023702} {\bibfield  {journal} {\bibinfo  {journal} {Phys. Rev. A}\ }\textbf {\bibinfo {volume} {102}},\ \bibinfo {pages} {023702} (\bibinfo {year} {2020})}\BibitemShut {NoStop}%
\bibitem [{\citenamefont {Le~Jeannic}\ \emph {et~al.}(2021)\citenamefont {Le~Jeannic}, \citenamefont {Ramos}, \citenamefont {Simonsen}, \citenamefont {Pregnolato}, \citenamefont {Liu}, \citenamefont {Schott}, \citenamefont {Wieck}, \citenamefont {Ludwig}, \citenamefont {Rotenberg}, \citenamefont {Garc\'{\i}a-Ripoll},\ and\ \citenamefont {Lodahl}}]{PhysRevLett.126.023603}%
  \BibitemOpen
  \bibfield  {author} {\bibinfo {author} {\bibfnamefont {H.}~\bibnamefont {Le~Jeannic}}, \bibinfo {author} {\bibfnamefont {T.}~\bibnamefont {Ramos}}, \bibinfo {author} {\bibfnamefont {S.~F.}\ \bibnamefont {Simonsen}}, \bibinfo {author} {\bibfnamefont {T.}~\bibnamefont {Pregnolato}}, \bibinfo {author} {\bibfnamefont {Z.}~\bibnamefont {Liu}}, \bibinfo {author} {\bibfnamefont {R.}~\bibnamefont {Schott}}, \bibinfo {author} {\bibfnamefont {A.~D.}\ \bibnamefont {Wieck}}, \bibinfo {author} {\bibfnamefont {A.}~\bibnamefont {Ludwig}}, \bibinfo {author} {\bibfnamefont {N.}~\bibnamefont {Rotenberg}}, \bibinfo {author} {\bibfnamefont {J.~J.}\ \bibnamefont {Garc\'{\i}a-Ripoll}},\ and\ \bibinfo {author} {\bibfnamefont {P.}~\bibnamefont {Lodahl}},\ }\bibfield  {title} {\bibinfo {title} {Experimental reconstruction of the few-photon nonlinear scattering matrix from a single quantum dot in a nanophotonic waveguide},\ }\href {https://doi.org/10.1103/PhysRevLett.126.023603} {\bibfield  {journal} {\bibinfo  {journal} {Phys. Rev.
  Lett.}\ }\textbf {\bibinfo {volume} {126}},\ \bibinfo {pages} {023603} (\bibinfo {year} {2021})}\BibitemShut {NoStop}%
\bibitem [{\citenamefont {Shen}\ and\ \citenamefont {Fan}(2007{\natexlab{a}})}]{PhysRevA.76.062709}%
  \BibitemOpen
  \bibfield  {author} {\bibinfo {author} {\bibfnamefont {J.-T.}\ \bibnamefont {Shen}}\ and\ \bibinfo {author} {\bibfnamefont {S.}~\bibnamefont {Fan}},\ }\bibfield  {title} {\bibinfo {title} {Strongly correlated multiparticle transport in one dimension through a quantum impurity},\ }\href {https://doi.org/10.1103/PhysRevA.76.062709} {\bibfield  {journal} {\bibinfo  {journal} {Phys. Rev. A}\ }\textbf {\bibinfo {volume} {76}},\ \bibinfo {pages} {062709} (\bibinfo {year} {2007}{\natexlab{a}})}\BibitemShut {NoStop}%
\bibitem [{\citenamefont {Shen}\ and\ \citenamefont {Fan}(2007{\natexlab{b}})}]{PhysRevLett.98.153003}%
  \BibitemOpen
  \bibfield  {author} {\bibinfo {author} {\bibfnamefont {J.-T.}\ \bibnamefont {Shen}}\ and\ \bibinfo {author} {\bibfnamefont {S.}~\bibnamefont {Fan}},\ }\bibfield  {title} {\bibinfo {title} {Strongly correlated two-photon transport in a one-dimensional waveguide coupled to a two-level system},\ }\href {https://doi.org/10.1103/PhysRevLett.98.153003} {\bibfield  {journal} {\bibinfo  {journal} {Phys. Rev. Lett.}\ }\textbf {\bibinfo {volume} {98}},\ \bibinfo {pages} {153003} (\bibinfo {year} {2007}{\natexlab{b}})}\BibitemShut {NoStop}%
\bibitem [{\citenamefont {Witthaut}\ and\ \citenamefont {S{\o}rensen}(2010)}]{Witthaut_2010}%
  \BibitemOpen
  \bibfield  {author} {\bibinfo {author} {\bibfnamefont {D.}~\bibnamefont {Witthaut}}\ and\ \bibinfo {author} {\bibfnamefont {A.~S.}\ \bibnamefont {S{\o}rensen}},\ }\bibfield  {title} {\bibinfo {title} {Photon scattering by a three-level emitter in a one-dimensional waveguide},\ }\href {https://doi.org/10.1088/1367-2630/12/4/043052} {\bibfield  {journal} {\bibinfo  {journal} {New Journal of Physics}\ }\textbf {\bibinfo {volume} {12}},\ \bibinfo {pages} {043052} (\bibinfo {year} {2010})}\BibitemShut {NoStop}%
\bibitem [{\citenamefont {S\'anchez-Burillo}\ \emph {et~al.}(2014)\citenamefont {S\'anchez-Burillo}, \citenamefont {Zueco}, \citenamefont {Garc\'ia-Ripoll},\ and\ \citenamefont {Mart\'in-Moreno}}]{PhysRevLett.113.263604}%
  \BibitemOpen
  \bibfield  {author} {\bibinfo {author} {\bibfnamefont {E.}~\bibnamefont {S\'anchez-Burillo}}, \bibinfo {author} {\bibfnamefont {D.}~\bibnamefont {Zueco}}, \bibinfo {author} {\bibfnamefont {J.~J.}\ \bibnamefont {Garc\'ia-Ripoll}},\ and\ \bibinfo {author} {\bibfnamefont {L.}~\bibnamefont {Mart\'in-Moreno}},\ }\bibfield  {title} {\bibinfo {title} {Scattering in the ultrastrong regime: Nonlinear optics with one photon},\ }\href {https://doi.org/10.1103/PhysRevLett.113.263604} {\bibfield  {journal} {\bibinfo  {journal} {Phys. Rev. Lett.}\ }\textbf {\bibinfo {volume} {113}},\ \bibinfo {pages} {263604} (\bibinfo {year} {2014})}\BibitemShut {NoStop}%
\bibitem [{\citenamefont {Calaj\'o}\ \emph {et~al.}(2016)\citenamefont {Calaj\'o}, \citenamefont {Ciccarello}, \citenamefont {Chang},\ and\ \citenamefont {Rabl}}]{Calaj2016}%
  \BibitemOpen
  \bibfield  {author} {\bibinfo {author} {\bibfnamefont {G.}~\bibnamefont {Calaj\'o}}, \bibinfo {author} {\bibfnamefont {F.}~\bibnamefont {Ciccarello}}, \bibinfo {author} {\bibfnamefont {D.}~\bibnamefont {Chang}},\ and\ \bibinfo {author} {\bibfnamefont {P.}~\bibnamefont {Rabl}},\ }\bibfield  {title} {\bibinfo {title} {Atom-field dressed states in slow-light waveguide {QED}},\ }\href {https://doi.org/10.1103/PhysRevA.93.033833} {\bibfield  {journal} {\bibinfo  {journal} {Phys. Rev. A}\ }\textbf {\bibinfo {volume} {93}},\ \bibinfo {pages} {033833} (\bibinfo {year} {2016})}\BibitemShut {NoStop}%
\bibitem [{\citenamefont {Pichler}\ and\ \citenamefont {Zoller}(2016)}]{PhysRevLett.116.093601}%
  \BibitemOpen
  \bibfield  {author} {\bibinfo {author} {\bibfnamefont {H.}~\bibnamefont {Pichler}}\ and\ \bibinfo {author} {\bibfnamefont {P.}~\bibnamefont {Zoller}},\ }\bibfield  {title} {\bibinfo {title} {Photonic circuits with time delays and quantum feedback},\ }\href {https://doi.org/10.1103/PhysRevLett.116.093601} {\bibfield  {journal} {\bibinfo  {journal} {Phys. Rev. Lett.}\ }\textbf {\bibinfo {volume} {116}},\ \bibinfo {pages} {093601} (\bibinfo {year} {2016})}\BibitemShut {NoStop}%
\bibitem [{\citenamefont {Young}\ \emph {et~al.}(2015)\citenamefont {Young}, \citenamefont {Thijssen}, \citenamefont {Beggs}, \citenamefont {Androvitsaneas}, \citenamefont {Kuipers}, \citenamefont {Rarity}, \citenamefont {Hughes},\ and\ \citenamefont {Oulton}}]{PhysRevLett.115.153901}%
  \BibitemOpen
  \bibfield  {author} {\bibinfo {author} {\bibfnamefont {A.~B.}\ \bibnamefont {Young}}, \bibinfo {author} {\bibfnamefont {A.~C.~T.}\ \bibnamefont {Thijssen}}, \bibinfo {author} {\bibfnamefont {D.~M.}\ \bibnamefont {Beggs}}, \bibinfo {author} {\bibfnamefont {P.}~\bibnamefont {Androvitsaneas}}, \bibinfo {author} {\bibfnamefont {L.}~\bibnamefont {Kuipers}}, \bibinfo {author} {\bibfnamefont {J.~G.}\ \bibnamefont {Rarity}}, \bibinfo {author} {\bibfnamefont {S.}~\bibnamefont {Hughes}},\ and\ \bibinfo {author} {\bibfnamefont {R.}~\bibnamefont {Oulton}},\ }\bibfield  {title} {\bibinfo {title} {Polarization engineering in photonic crystal waveguides for spin-photon entanglers},\ }\href {https://doi.org/10.1103/PhysRevLett.115.153901} {\bibfield  {journal} {\bibinfo  {journal} {Phys. Rev. Lett.}\ }\textbf {\bibinfo {volume} {115}},\ \bibinfo {pages} {153901} (\bibinfo {year} {2015})}\BibitemShut {NoStop}%
\bibitem [{\citenamefont {Lodahl}\ \emph {et~al.}(2017)\citenamefont {Lodahl}, \citenamefont {Mahmoodian}, \citenamefont {Stobbe}, \citenamefont {Rauschenbeutel}, \citenamefont {Schneeweiss}, \citenamefont {Volz}, \citenamefont {Pichler},\ and\ \citenamefont {Zoller}}]{Lodahl2017}%
  \BibitemOpen
  \bibfield  {author} {\bibinfo {author} {\bibfnamefont {P.}~\bibnamefont {Lodahl}}, \bibinfo {author} {\bibfnamefont {S.}~\bibnamefont {Mahmoodian}}, \bibinfo {author} {\bibfnamefont {S.}~\bibnamefont {Stobbe}}, \bibinfo {author} {\bibfnamefont {A.}~\bibnamefont {Rauschenbeutel}}, \bibinfo {author} {\bibfnamefont {P.}~\bibnamefont {Schneeweiss}}, \bibinfo {author} {\bibfnamefont {J.}~\bibnamefont {Volz}}, \bibinfo {author} {\bibfnamefont {H.}~\bibnamefont {Pichler}},\ and\ \bibinfo {author} {\bibfnamefont {P.}~\bibnamefont {Zoller}},\ }\bibfield  {title} {\bibinfo {title} {Chiral quantum optics},\ }\href {https://doi.org/10.1038/nature21037} {\bibfield  {journal} {\bibinfo  {journal} {Nature}\ }\textbf {\bibinfo {volume} {541}},\ \bibinfo {pages} {473} (\bibinfo {year} {2017})}\BibitemShut {NoStop}%
\bibitem [{\citenamefont {Hauff}\ \emph {et~al.}(2022)\citenamefont {Hauff}, \citenamefont {Le~Jeannic}, \citenamefont {Lodahl}, \citenamefont {Hughes},\ and\ \citenamefont {Rotenberg}}]{PhysRevResearch.4.023082}%
  \BibitemOpen
  \bibfield  {author} {\bibinfo {author} {\bibfnamefont {N.~V.}\ \bibnamefont {Hauff}}, \bibinfo {author} {\bibfnamefont {H.}~\bibnamefont {Le~Jeannic}}, \bibinfo {author} {\bibfnamefont {P.}~\bibnamefont {Lodahl}}, \bibinfo {author} {\bibfnamefont {S.}~\bibnamefont {Hughes}},\ and\ \bibinfo {author} {\bibfnamefont {N.}~\bibnamefont {Rotenberg}},\ }\bibfield  {title} {\bibinfo {title} {Chiral quantum optics in broken-symmetry and topological photonic crystal waveguides},\ }\href {https://doi.org/10.1103/PhysRevResearch.4.023082} {\bibfield  {journal} {\bibinfo  {journal} {Phys. Rev. Res.}\ }\textbf {\bibinfo {volume} {4}},\ \bibinfo {pages} {023082} (\bibinfo {year} {2022})}\BibitemShut {NoStop}%
\bibitem [{\citenamefont {Mirhosseini}\ \emph {et~al.}(2019)\citenamefont {Mirhosseini}, \citenamefont {Kim}, \citenamefont {Zhang}, \citenamefont {Sipahigil}, \citenamefont {Dieterle}, \citenamefont {Keller}, \citenamefont {Asenjo-Garcia}, \citenamefont {Chang},\ and\ \citenamefont {Painter}}]{Mirhosseini2019}%
  \BibitemOpen
  \bibfield  {author} {\bibinfo {author} {\bibfnamefont {M.}~\bibnamefont {Mirhosseini}}, \bibinfo {author} {\bibfnamefont {E.}~\bibnamefont {Kim}}, \bibinfo {author} {\bibfnamefont {X.}~\bibnamefont {Zhang}}, \bibinfo {author} {\bibfnamefont {A.}~\bibnamefont {Sipahigil}}, \bibinfo {author} {\bibfnamefont {P.~B.}\ \bibnamefont {Dieterle}}, \bibinfo {author} {\bibfnamefont {A.~J.}\ \bibnamefont {Keller}}, \bibinfo {author} {\bibfnamefont {A.}~\bibnamefont {Asenjo-Garcia}}, \bibinfo {author} {\bibfnamefont {D.~E.}\ \bibnamefont {Chang}},\ and\ \bibinfo {author} {\bibfnamefont {O.}~\bibnamefont {Painter}},\ }\bibfield  {title} {\bibinfo {title} {Cavity quantum electrodynamics with atom-like mirrors},\ }\href {https://doi.org/10.1038/s41586-019-1196-1} {\bibfield  {journal} {\bibinfo  {journal} {Nature}\ }\textbf {\bibinfo {volume} {569}},\ \bibinfo {pages} {692} (\bibinfo {year} {2019})}\BibitemShut {NoStop}%
\bibitem [{\citenamefont {T\"{u}rschmann}\ \emph {et~al.}(2019)\citenamefont {T\"{u}rschmann}, \citenamefont {Jeannic}, \citenamefont {Simonsen}, \citenamefont {Haakh}, \citenamefont {G\"{o}tzinger}, \citenamefont {Sandoghdar}, \citenamefont {Lodahl},\ and\ \citenamefont {Rotenberg}}]{Trschmann2019}%
  \BibitemOpen
  \bibfield  {author} {\bibinfo {author} {\bibfnamefont {P.}~\bibnamefont {T\"{u}rschmann}}, \bibinfo {author} {\bibfnamefont {H.~L.}\ \bibnamefont {Jeannic}}, \bibinfo {author} {\bibfnamefont {S.~F.}\ \bibnamefont {Simonsen}}, \bibinfo {author} {\bibfnamefont {H.~R.}\ \bibnamefont {Haakh}}, \bibinfo {author} {\bibfnamefont {S.}~\bibnamefont {G\"{o}tzinger}}, \bibinfo {author} {\bibfnamefont {V.}~\bibnamefont {Sandoghdar}}, \bibinfo {author} {\bibfnamefont {P.}~\bibnamefont {Lodahl}},\ and\ \bibinfo {author} {\bibfnamefont {N.}~\bibnamefont {Rotenberg}},\ }\bibfield  {title} {\bibinfo {title} {Coherent nonlinear optics of quantum emitters in nanophotonic waveguides},\ }\href {https://doi.org/10.1515/nanoph-2019-0126} {\bibfield  {journal} {\bibinfo  {journal} {Nanophotonics}\ }\textbf {\bibinfo {volume} {8}},\ \bibinfo {pages} {1641} (\bibinfo {year} {2019})}\BibitemShut {NoStop}%
\bibitem [{\citenamefont {Pregnolato}\ \emph {et~al.}(2020)\citenamefont {Pregnolato}, \citenamefont {Chu}, \citenamefont {Schr\"oder}, \citenamefont {Schott}, \citenamefont {Wieck}, \citenamefont {Ludwig}, \citenamefont {Lodahl},\ and\ \citenamefont {Rotenberg}}]{10.1063/1.5117888}%
  \BibitemOpen
  \bibfield  {author} {\bibinfo {author} {\bibfnamefont {T.}~\bibnamefont {Pregnolato}}, \bibinfo {author} {\bibfnamefont {X.-L.}\ \bibnamefont {Chu}}, \bibinfo {author} {\bibfnamefont {T.}~\bibnamefont {Schr\"oder}}, \bibinfo {author} {\bibfnamefont {R.}~\bibnamefont {Schott}}, \bibinfo {author} {\bibfnamefont {A.~D.}\ \bibnamefont {Wieck}}, \bibinfo {author} {\bibfnamefont {A.}~\bibnamefont {Ludwig}}, \bibinfo {author} {\bibfnamefont {P.}~\bibnamefont {Lodahl}},\ and\ \bibinfo {author} {\bibfnamefont {N.}~\bibnamefont {Rotenberg}},\ }\bibfield  {title} {\bibinfo {title} {{Deterministic positioning of nanophotonic waveguides around single self-assembled quantum dots}},\ }\href {https://doi.org/10.1063/1.5117888} {\bibfield  {journal} {\bibinfo  {journal} {APL Photonics}\ }\textbf {\bibinfo {volume} {5}},\ \bibinfo {pages} {086101} (\bibinfo {year} {2020})}\BibitemShut {NoStop}%
\bibitem [{\citenamefont {Lund-Hansen}\ \emph {et~al.}(2008)\citenamefont {Lund-Hansen}, \citenamefont {Stobbe}, \citenamefont {Julsgaard}, \citenamefont {Thyrrestrup}, \citenamefont {S\"unner}, \citenamefont {Kamp}, \citenamefont {Forchel},\ and\ \citenamefont {Lodahl}}]{PhysRevLett.101.113903}%
  \BibitemOpen
  \bibfield  {author} {\bibinfo {author} {\bibfnamefont {T.}~\bibnamefont {Lund-Hansen}}, \bibinfo {author} {\bibfnamefont {S.}~\bibnamefont {Stobbe}}, \bibinfo {author} {\bibfnamefont {B.}~\bibnamefont {Julsgaard}}, \bibinfo {author} {\bibfnamefont {H.}~\bibnamefont {Thyrrestrup}}, \bibinfo {author} {\bibfnamefont {T.}~\bibnamefont {S\"unner}}, \bibinfo {author} {\bibfnamefont {M.}~\bibnamefont {Kamp}}, \bibinfo {author} {\bibfnamefont {A.}~\bibnamefont {Forchel}},\ and\ \bibinfo {author} {\bibfnamefont {P.}~\bibnamefont {Lodahl}},\ }\bibfield  {title} {\bibinfo {title} {Experimental realization of highly efficient broadband coupling of single quantum dots to a photonic crystal waveguide},\ }\href {https://doi.org/10.1103/PhysRevLett.101.113903} {\bibfield  {journal} {\bibinfo  {journal} {Phys. Rev. Lett.}\ }\textbf {\bibinfo {volume} {101}},\ \bibinfo {pages} {113903} (\bibinfo {year} {2008})}\BibitemShut {NoStop}%
\bibitem [{\citenamefont {Gonz\'alez-Tudela}\ \emph {et~al.}(2015)\citenamefont {Gonz\'alez-Tudela}, \citenamefont {Paulisch}, \citenamefont {Chang}, \citenamefont {Kimble},\ and\ \citenamefont {Cirac}}]{PhysRevLett.115.163603}%
  \BibitemOpen
  \bibfield  {author} {\bibinfo {author} {\bibfnamefont {A.}~\bibnamefont {Gonz\'alez-Tudela}}, \bibinfo {author} {\bibfnamefont {V.}~\bibnamefont {Paulisch}}, \bibinfo {author} {\bibfnamefont {D.~E.}\ \bibnamefont {Chang}}, \bibinfo {author} {\bibfnamefont {H.~J.}\ \bibnamefont {Kimble}},\ and\ \bibinfo {author} {\bibfnamefont {J.~I.}\ \bibnamefont {Cirac}},\ }\bibfield  {title} {\bibinfo {title} {Deterministic generation of arbitrary photonic states assisted by dissipation},\ }\href {https://doi.org/10.1103/PhysRevLett.115.163603} {\bibfield  {journal} {\bibinfo  {journal} {Phys. Rev. Lett.}\ }\textbf {\bibinfo {volume} {115}},\ \bibinfo {pages} {163603} (\bibinfo {year} {2015})}\BibitemShut {NoStop}%
\bibitem [{\citenamefont {Laucht}\ \emph {et~al.}(2012)\citenamefont {Laucht}, \citenamefont {P\"utz}, \citenamefont {G\"unthner}, \citenamefont {Hauke}, \citenamefont {Saive}, \citenamefont {Fr\'ed\'erick}, \citenamefont {Bichler}, \citenamefont {Amann}, \citenamefont {Holleitner}, \citenamefont {Kaniber},\ and\ \citenamefont {Finley}}]{PhysRevX.2.011014}%
  \BibitemOpen
  \bibfield  {author} {\bibinfo {author} {\bibfnamefont {A.}~\bibnamefont {Laucht}}, \bibinfo {author} {\bibfnamefont {S.}~\bibnamefont {P\"utz}}, \bibinfo {author} {\bibfnamefont {T.}~\bibnamefont {G\"unthner}}, \bibinfo {author} {\bibfnamefont {N.}~\bibnamefont {Hauke}}, \bibinfo {author} {\bibfnamefont {R.}~\bibnamefont {Saive}}, \bibinfo {author} {\bibfnamefont {S.}~\bibnamefont {Fr\'ed\'erick}}, \bibinfo {author} {\bibfnamefont {M.}~\bibnamefont {Bichler}}, \bibinfo {author} {\bibfnamefont {M.-C.}\ \bibnamefont {Amann}}, \bibinfo {author} {\bibfnamefont {A.~W.}\ \bibnamefont {Holleitner}}, \bibinfo {author} {\bibfnamefont {M.}~\bibnamefont {Kaniber}},\ and\ \bibinfo {author} {\bibfnamefont {J.~J.}\ \bibnamefont {Finley}},\ }\bibfield  {title} {\bibinfo {title} {A waveguide-coupled on-chip single-photon source},\ }\href {https://doi.org/10.1103/PhysRevX.2.011014} {\bibfield  {journal} {\bibinfo  {journal} {Phys. Rev. X}\ }\textbf {\bibinfo {volume} {2}},\ \bibinfo {pages} {011014} (\bibinfo {year}
  {2012})}\BibitemShut {NoStop}%
\bibitem [{\citenamefont {Nie}\ \emph {et~al.}(2023)\citenamefont {Nie}, \citenamefont {Shi}, \citenamefont {Liu},\ and\ \citenamefont {Nori}}]{PhysRevLett.131.103602}%
  \BibitemOpen
  \bibfield  {author} {\bibinfo {author} {\bibfnamefont {W.}~\bibnamefont {Nie}}, \bibinfo {author} {\bibfnamefont {T.}~\bibnamefont {Shi}}, \bibinfo {author} {\bibfnamefont {Y.-x.}\ \bibnamefont {Liu}},\ and\ \bibinfo {author} {\bibfnamefont {F.}~\bibnamefont {Nori}},\ }\bibfield  {title} {\bibinfo {title} {{Non-Hermitian} waveguide cavity {QED} with tunable atomic mirrors},\ }\href {https://doi.org/10.1103/PhysRevLett.131.103602} {\bibfield  {journal} {\bibinfo  {journal} {Phys. Rev. Lett.}\ }\textbf {\bibinfo {volume} {131}},\ \bibinfo {pages} {103602} (\bibinfo {year} {2023})}\BibitemShut {NoStop}%
\bibitem [{\citenamefont {Li}\ and\ \citenamefont {Wei}(2015)}]{PhysRevA.92.063836}%
  \BibitemOpen
  \bibfield  {author} {\bibinfo {author} {\bibfnamefont {X.}~\bibnamefont {Li}}\ and\ \bibinfo {author} {\bibfnamefont {L.~F.}\ \bibnamefont {Wei}},\ }\bibfield  {title} {\bibinfo {title} {Designable single-photon quantum routings with atomic mirrors},\ }\href {https://doi.org/10.1103/PhysRevA.92.063836} {\bibfield  {journal} {\bibinfo  {journal} {Phys. Rev. A}\ }\textbf {\bibinfo {volume} {92}},\ \bibinfo {pages} {063836} (\bibinfo {year} {2015})}\BibitemShut {NoStop}%
\bibitem [{\citenamefont {Asenjo-Garcia}\ \emph {et~al.}(2017)\citenamefont {Asenjo-Garcia}, \citenamefont {Moreno-Cardoner}, \citenamefont {Albrecht}, \citenamefont {Kimble},\ and\ \citenamefont {Chang}}]{PhysRevX.7.031024}%
  \BibitemOpen
  \bibfield  {author} {\bibinfo {author} {\bibfnamefont {A.}~\bibnamefont {Asenjo-Garcia}}, \bibinfo {author} {\bibfnamefont {M.}~\bibnamefont {Moreno-Cardoner}}, \bibinfo {author} {\bibfnamefont {A.}~\bibnamefont {Albrecht}}, \bibinfo {author} {\bibfnamefont {H.~J.}\ \bibnamefont {Kimble}},\ and\ \bibinfo {author} {\bibfnamefont {D.~E.}\ \bibnamefont {Chang}},\ }\bibfield  {title} {\bibinfo {title} {Exponential improvement in photon storage fidelities using subradiance and ``selective radiance'' in atomic arrays},\ }\href {https://doi.org/10.1103/PhysRevX.7.031024} {\bibfield  {journal} {\bibinfo  {journal} {Phys. Rev. X}\ }\textbf {\bibinfo {volume} {7}},\ \bibinfo {pages} {031024} (\bibinfo {year} {2017})}\BibitemShut {NoStop}%
\bibitem [{\citenamefont {Kockum}\ \emph {et~al.}(2018)\citenamefont {Kockum}, \citenamefont {Johansson},\ and\ \citenamefont {Nori}}]{PhysRevLett.120.140404}%
  \BibitemOpen
  \bibfield  {author} {\bibinfo {author} {\bibfnamefont {A.~F.}\ \bibnamefont {Kockum}}, \bibinfo {author} {\bibfnamefont {G.}~\bibnamefont {Johansson}},\ and\ \bibinfo {author} {\bibfnamefont {F.}~\bibnamefont {Nori}},\ }\bibfield  {title} {\bibinfo {title} {Decoherence-free interaction between giant atoms in waveguide quantum electrodynamics},\ }\href {https://doi.org/10.1103/PhysRevLett.120.140404} {\bibfield  {journal} {\bibinfo  {journal} {Phys. Rev. Lett.}\ }\textbf {\bibinfo {volume} {120}},\ \bibinfo {pages} {140404} (\bibinfo {year} {2018})}\BibitemShut {NoStop}%
\bibitem [{\citenamefont {Chu}\ \emph {et~al.}(2023)\citenamefont {Chu}, \citenamefont {Papon}, \citenamefont {Bart}, \citenamefont {Wieck}, \citenamefont {Ludwig}, \citenamefont {Midolo}, \citenamefont {Rotenberg},\ and\ \citenamefont {Lodahl}}]{PhysRevLett.131.033606}%
  \BibitemOpen
  \bibfield  {author} {\bibinfo {author} {\bibfnamefont {X.-L.}\ \bibnamefont {Chu}}, \bibinfo {author} {\bibfnamefont {C.}~\bibnamefont {Papon}}, \bibinfo {author} {\bibfnamefont {N.}~\bibnamefont {Bart}}, \bibinfo {author} {\bibfnamefont {A.~D.}\ \bibnamefont {Wieck}}, \bibinfo {author} {\bibfnamefont {A.}~\bibnamefont {Ludwig}}, \bibinfo {author} {\bibfnamefont {L.}~\bibnamefont {Midolo}}, \bibinfo {author} {\bibfnamefont {N.}~\bibnamefont {Rotenberg}},\ and\ \bibinfo {author} {\bibfnamefont {P.}~\bibnamefont {Lodahl}},\ }\bibfield  {title} {\bibinfo {title} {Independent electrical control of two quantum dots coupled through a photonic-crystal waveguide},\ }\href {https://doi.org/10.1103/PhysRevLett.131.033606} {\bibfield  {journal} {\bibinfo  {journal} {Phys. Rev. Lett.}\ }\textbf {\bibinfo {volume} {131}},\ \bibinfo {pages} {033606} (\bibinfo {year} {2023})}\BibitemShut {NoStop}%
\bibitem [{\citenamefont {Masson}\ and\ \citenamefont {Asenjo-Garcia}(2020)}]{PhysRevResearch.2.043213}%
  \BibitemOpen
  \bibfield  {author} {\bibinfo {author} {\bibfnamefont {S.~J.}\ \bibnamefont {Masson}}\ and\ \bibinfo {author} {\bibfnamefont {A.}~\bibnamefont {Asenjo-Garcia}},\ }\bibfield  {title} {\bibinfo {title} {Atomic-waveguide quantum electrodynamics},\ }\href {https://doi.org/10.1103/PhysRevResearch.2.043213} {\bibfield  {journal} {\bibinfo  {journal} {Phys. Rev. Res.}\ }\textbf {\bibinfo {volume} {2}},\ \bibinfo {pages} {043213} (\bibinfo {year} {2020})}\BibitemShut {NoStop}%
\bibitem [{\citenamefont {Kannan}\ \emph {et~al.}(2020)\citenamefont {Kannan}, \citenamefont {Ruckriegel}, \citenamefont {Campbell}, \citenamefont {Frisk~Kockum}, \citenamefont {Braum\"{u}ller}, \citenamefont {Kim}, \citenamefont {Kjaergaard}, \citenamefont {Krantz}, \citenamefont {Melville}, \citenamefont {Niedzielski}, \citenamefont {Veps\"{a}l\"{a}inen}, \citenamefont {Winik}, \citenamefont {Yoder}, \citenamefont {Nori}, \citenamefont {Orlando}, \citenamefont {Gustavsson},\ and\ \citenamefont {Oliver}}]{Kannan2020}%
  \BibitemOpen
  \bibfield  {author} {\bibinfo {author} {\bibfnamefont {B.}~\bibnamefont {Kannan}}, \bibinfo {author} {\bibfnamefont {M.~J.}\ \bibnamefont {Ruckriegel}}, \bibinfo {author} {\bibfnamefont {D.~L.}\ \bibnamefont {Campbell}}, \bibinfo {author} {\bibfnamefont {A.}~\bibnamefont {Frisk~Kockum}}, \bibinfo {author} {\bibfnamefont {J.}~\bibnamefont {Braum\"{u}ller}}, \bibinfo {author} {\bibfnamefont {D.~K.}\ \bibnamefont {Kim}}, \bibinfo {author} {\bibfnamefont {M.}~\bibnamefont {Kjaergaard}}, \bibinfo {author} {\bibfnamefont {P.}~\bibnamefont {Krantz}}, \bibinfo {author} {\bibfnamefont {A.}~\bibnamefont {Melville}}, \bibinfo {author} {\bibfnamefont {B.~M.}\ \bibnamefont {Niedzielski}}, \bibinfo {author} {\bibfnamefont {A.}~\bibnamefont {Veps\"{a}l\"{a}inen}}, \bibinfo {author} {\bibfnamefont {R.}~\bibnamefont {Winik}}, \bibinfo {author} {\bibfnamefont {J.~L.}\ \bibnamefont {Yoder}}, \bibinfo {author} {\bibfnamefont {F.}~\bibnamefont {Nori}}, \bibinfo {author} {\bibfnamefont {T.~P.}\ \bibnamefont {Orlando}}, \bibinfo
  {author} {\bibfnamefont {S.}~\bibnamefont {Gustavsson}},\ and\ \bibinfo {author} {\bibfnamefont {W.~D.}\ \bibnamefont {Oliver}},\ }\bibfield  {title} {\bibinfo {title} {Waveguide quantum electrodynamics with superconducting artificial giant atoms},\ }\href {https://doi.org/10.1038/s41586-020-2529-9} {\bibfield  {journal} {\bibinfo  {journal} {Nature}\ }\textbf {\bibinfo {volume} {583}},\ \bibinfo {pages} {775} (\bibinfo {year} {2020})}\BibitemShut {NoStop}%
\bibitem [{\citenamefont {Alushi}\ \emph {et~al.}(2023)\citenamefont {Alushi}, \citenamefont {Ramos}, \citenamefont {Garc\'{\i}a-Ripoll}, \citenamefont {Di~Candia},\ and\ \citenamefont {Felicetti}}]{PRXQuantum.4.030326}%
  \BibitemOpen
  \bibfield  {author} {\bibinfo {author} {\bibfnamefont {U.}~\bibnamefont {Alushi}}, \bibinfo {author} {\bibfnamefont {T.}~\bibnamefont {Ramos}}, \bibinfo {author} {\bibfnamefont {J.~J.}\ \bibnamefont {Garc\'{\i}a-Ripoll}}, \bibinfo {author} {\bibfnamefont {R.}~\bibnamefont {Di~Candia}},\ and\ \bibinfo {author} {\bibfnamefont {S.}~\bibnamefont {Felicetti}},\ }\bibfield  {title} {\bibinfo {title} {Waveguide {QED} with quadratic light-matter interactions},\ }\href {https://doi.org/10.1103/PRXQuantum.4.030326} {\bibfield  {journal} {\bibinfo  {journal} {PRX Quantum}\ }\textbf {\bibinfo {volume} {4}},\ \bibinfo {pages} {030326} (\bibinfo {year} {2023})}\BibitemShut {NoStop}%
\bibitem [{\citenamefont {Wang}\ \emph {et~al.}(2024)\citenamefont {Wang}, \citenamefont {Zhu}, \citenamefont {Liu},\ and\ \citenamefont {Nori}}]{PhysRevResearch.6.013279}%
  \BibitemOpen
  \bibfield  {author} {\bibinfo {author} {\bibfnamefont {X.}~\bibnamefont {Wang}}, \bibinfo {author} {\bibfnamefont {H.-B.}\ \bibnamefont {Zhu}}, \bibinfo {author} {\bibfnamefont {T.}~\bibnamefont {Liu}},\ and\ \bibinfo {author} {\bibfnamefont {F.}~\bibnamefont {Nori}},\ }\bibfield  {title} {\bibinfo {title} {Realizing quantum optics in structured environments with giant atoms},\ }\href {https://doi.org/10.1103/PhysRevResearch.6.013279} {\bibfield  {journal} {\bibinfo  {journal} {Phys. Rev. Res.}\ }\textbf {\bibinfo {volume} {6}},\ \bibinfo {pages} {013279} (\bibinfo {year} {2024})}\BibitemShut {NoStop}%
\bibitem [{\citenamefont {Wang}\ and\ \citenamefont {Li}(2022)}]{Wang_2022}%
  \BibitemOpen
  \bibfield  {author} {\bibinfo {author} {\bibfnamefont {X.}~\bibnamefont {Wang}}\ and\ \bibinfo {author} {\bibfnamefont {H.-R.}\ \bibnamefont {Li}},\ }\bibfield  {title} {\bibinfo {title} {Chiral quantum network with giant atoms},\ }\href {https://doi.org/10.1088/2058-9565/ac6a04} {\bibfield  {journal} {\bibinfo  {journal} {Quantum Science and Technology}\ }\textbf {\bibinfo {volume} {7}},\ \bibinfo {pages} {035007} (\bibinfo {year} {2022})}\BibitemShut {NoStop}%
\bibitem [{\citenamefont {Astafiev}\ \emph {et~al.}(2010)\citenamefont {Astafiev}, \citenamefont {Zagoskin}, \citenamefont {Abdumalikov}, \citenamefont {Pashkin}, \citenamefont {Yamamoto}, \citenamefont {Inomata}, \citenamefont {Nakamura},\ and\ \citenamefont {Tsai}}]{Astafiev2010}%
  \BibitemOpen
  \bibfield  {author} {\bibinfo {author} {\bibfnamefont {O.}~\bibnamefont {Astafiev}}, \bibinfo {author} {\bibfnamefont {A.~M.}\ \bibnamefont {Zagoskin}}, \bibinfo {author} {\bibfnamefont {A.~A.}\ \bibnamefont {Abdumalikov}}, \bibinfo {author} {\bibfnamefont {Y.~A.}\ \bibnamefont {Pashkin}}, \bibinfo {author} {\bibfnamefont {T.}~\bibnamefont {Yamamoto}}, \bibinfo {author} {\bibfnamefont {K.}~\bibnamefont {Inomata}}, \bibinfo {author} {\bibfnamefont {Y.}~\bibnamefont {Nakamura}},\ and\ \bibinfo {author} {\bibfnamefont {J.~S.}\ \bibnamefont {Tsai}},\ }\bibfield  {title} {\bibinfo {title} {Resonance fluorescence of a single artificial atom},\ }\href {https://doi.org/10.1126/science.1181918} {\bibfield  {journal} {\bibinfo  {journal} {Science}\ }\textbf {\bibinfo {volume} {327}},\ \bibinfo {pages} {840–843} (\bibinfo {year} {2010})}\BibitemShut {NoStop}%
\bibitem [{\citenamefont {Hoi}\ \emph {et~al.}(2012)\citenamefont {Hoi}, \citenamefont {Palomaki}, \citenamefont {Lindkvist}, \citenamefont {Johansson}, \citenamefont {Delsing},\ and\ \citenamefont {Wilson}}]{PhysRevLett.108.263601}%
  \BibitemOpen
  \bibfield  {author} {\bibinfo {author} {\bibfnamefont {I.-C.}\ \bibnamefont {Hoi}}, \bibinfo {author} {\bibfnamefont {T.}~\bibnamefont {Palomaki}}, \bibinfo {author} {\bibfnamefont {J.}~\bibnamefont {Lindkvist}}, \bibinfo {author} {\bibfnamefont {G.}~\bibnamefont {Johansson}}, \bibinfo {author} {\bibfnamefont {P.}~\bibnamefont {Delsing}},\ and\ \bibinfo {author} {\bibfnamefont {C.~M.}\ \bibnamefont {Wilson}},\ }\bibfield  {title} {\bibinfo {title} {Generation of nonclassical microwave states using an artificial atom in 1d open space},\ }\href {https://doi.org/10.1103/PhysRevLett.108.263601} {\bibfield  {journal} {\bibinfo  {journal} {Phys. Rev. Lett.}\ }\textbf {\bibinfo {volume} {108}},\ \bibinfo {pages} {263601} (\bibinfo {year} {2012})}\BibitemShut {NoStop}%
\bibitem [{\citenamefont {Manga~Rao}\ and\ \citenamefont {Hughes}(2007)}]{PhysRevB.75.205437}%
  \BibitemOpen
  \bibfield  {author} {\bibinfo {author} {\bibfnamefont {V.~S.~C.}\ \bibnamefont {Manga~Rao}}\ and\ \bibinfo {author} {\bibfnamefont {S.}~\bibnamefont {Hughes}},\ }\bibfield  {title} {\bibinfo {title} {Single quantum-dot {Purcell} factor and $\ensuremath{\beta}$ factor in a photonic crystal waveguide},\ }\href {https://doi.org/10.1103/PhysRevB.75.205437} {\bibfield  {journal} {\bibinfo  {journal} {Phys. Rev. B}\ }\textbf {\bibinfo {volume} {75}},\ \bibinfo {pages} {205437} (\bibinfo {year} {2007})}\BibitemShut {NoStop}%
\bibitem [{\citenamefont {Arcari}\ \emph {et~al.}(2014)\citenamefont {Arcari}, \citenamefont {S\"ollner}, \citenamefont {Javadi}, \citenamefont {Lindskov~Hansen}, \citenamefont {Mahmoodian}, \citenamefont {Liu}, \citenamefont {Thyrrestrup}, \citenamefont {Lee}, \citenamefont {Song}, \citenamefont {Stobbe},\ and\ \citenamefont {Lodahl}}]{PhysRevLett.113.093603}%
  \BibitemOpen
  \bibfield  {author} {\bibinfo {author} {\bibfnamefont {M.}~\bibnamefont {Arcari}}, \bibinfo {author} {\bibfnamefont {I.}~\bibnamefont {S\"ollner}}, \bibinfo {author} {\bibfnamefont {A.}~\bibnamefont {Javadi}}, \bibinfo {author} {\bibfnamefont {S.}~\bibnamefont {Lindskov~Hansen}}, \bibinfo {author} {\bibfnamefont {S.}~\bibnamefont {Mahmoodian}}, \bibinfo {author} {\bibfnamefont {J.}~\bibnamefont {Liu}}, \bibinfo {author} {\bibfnamefont {H.}~\bibnamefont {Thyrrestrup}}, \bibinfo {author} {\bibfnamefont {E.~H.}\ \bibnamefont {Lee}}, \bibinfo {author} {\bibfnamefont {J.~D.}\ \bibnamefont {Song}}, \bibinfo {author} {\bibfnamefont {S.}~\bibnamefont {Stobbe}},\ and\ \bibinfo {author} {\bibfnamefont {P.}~\bibnamefont {Lodahl}},\ }\bibfield  {title} {\bibinfo {title} {Near-unity coupling efficiency of a quantum emitter to a photonic crystal waveguide},\ }\href {https://doi.org/10.1103/PhysRevLett.113.093603} {\bibfield  {journal} {\bibinfo  {journal} {Phys. Rev. Lett.}\ }\textbf {\bibinfo {volume} {113}},\ \bibinfo
  {pages} {093603} (\bibinfo {year} {2014})}\BibitemShut {NoStop}%
\bibitem [{\citenamefont {Paesani}\ \emph {et~al.}(2019)\citenamefont {Paesani}, \citenamefont {Ding}, \citenamefont {Santagati}, \citenamefont {Chakhmakhchyan}, \citenamefont {Vigliar}, \citenamefont {Rottwitt}, \citenamefont {Oxenl{\o}we}, \citenamefont {Wang}, \citenamefont {Thompson},\ and\ \citenamefont {Laing}}]{Paesani2019}%
  \BibitemOpen
  \bibfield  {author} {\bibinfo {author} {\bibfnamefont {S.}~\bibnamefont {Paesani}}, \bibinfo {author} {\bibfnamefont {Y.}~\bibnamefont {Ding}}, \bibinfo {author} {\bibfnamefont {R.}~\bibnamefont {Santagati}}, \bibinfo {author} {\bibfnamefont {L.}~\bibnamefont {Chakhmakhchyan}}, \bibinfo {author} {\bibfnamefont {C.}~\bibnamefont {Vigliar}}, \bibinfo {author} {\bibfnamefont {K.}~\bibnamefont {Rottwitt}}, \bibinfo {author} {\bibfnamefont {L.~K.}\ \bibnamefont {Oxenl{\o}we}}, \bibinfo {author} {\bibfnamefont {J.}~\bibnamefont {Wang}}, \bibinfo {author} {\bibfnamefont {M.~G.}\ \bibnamefont {Thompson}},\ and\ \bibinfo {author} {\bibfnamefont {A.}~\bibnamefont {Laing}},\ }\bibfield  {title} {\bibinfo {title} {Generation and sampling of quantum states of light in a silicon chip},\ }\href {https://doi.org/10.1038/s41567-019-0567-8} {\bibfield  {journal} {\bibinfo  {journal} {Nature Physics}\ }\textbf {\bibinfo {volume} {15}},\ \bibinfo {pages} {925} (\bibinfo {year} {2019})}\BibitemShut {NoStop}%
\bibitem [{\citenamefont {le~Feber}\ \emph {et~al.}(2015)\citenamefont {le~Feber}, \citenamefont {Rotenberg},\ and\ \citenamefont {Kuipers}}]{leFeber2015}%
  \BibitemOpen
  \bibfield  {author} {\bibinfo {author} {\bibfnamefont {B.}~\bibnamefont {le~Feber}}, \bibinfo {author} {\bibfnamefont {N.}~\bibnamefont {Rotenberg}},\ and\ \bibinfo {author} {\bibfnamefont {L.}~\bibnamefont {Kuipers}},\ }\bibfield  {title} {\bibinfo {title} {Nanophotonic control of circular dipole emission},\ }\bibfield  {journal} {\bibinfo  {journal} {Nature Communications}\ }\textbf {\bibinfo {volume} {6}},\ \href {https://doi.org/10.1038/ncomms7695} {10.1038/ncomms7695} (\bibinfo {year} {2015})\BibitemShut {NoStop}%
\bibitem [{\citenamefont {S\"{o}llner}\ \emph {et~al.}(2015)\citenamefont {S\"{o}llner}, \citenamefont {Mahmoodian}, \citenamefont {Hansen}, \citenamefont {Midolo}, \citenamefont {Javadi}, \citenamefont {Kir{\v{s}}ansk{\.{e}}}, \citenamefont {Pregnolato}, \citenamefont {El-Ella}, \citenamefont {Lee}, \citenamefont {Song}, \citenamefont {Stobbe},\ and\ \citenamefont {Lodahl}}]{Sllner2015}%
  \BibitemOpen
  \bibfield  {author} {\bibinfo {author} {\bibfnamefont {I.}~\bibnamefont {S\"{o}llner}}, \bibinfo {author} {\bibfnamefont {S.}~\bibnamefont {Mahmoodian}}, \bibinfo {author} {\bibfnamefont {S.~L.}\ \bibnamefont {Hansen}}, \bibinfo {author} {\bibfnamefont {L.}~\bibnamefont {Midolo}}, \bibinfo {author} {\bibfnamefont {A.}~\bibnamefont {Javadi}}, \bibinfo {author} {\bibfnamefont {G.}~\bibnamefont {Kir{\v{s}}ansk{\.{e}}}}, \bibinfo {author} {\bibfnamefont {T.}~\bibnamefont {Pregnolato}}, \bibinfo {author} {\bibfnamefont {H.}~\bibnamefont {El-Ella}}, \bibinfo {author} {\bibfnamefont {E.~H.}\ \bibnamefont {Lee}}, \bibinfo {author} {\bibfnamefont {J.~D.}\ \bibnamefont {Song}}, \bibinfo {author} {\bibfnamefont {S.}~\bibnamefont {Stobbe}},\ and\ \bibinfo {author} {\bibfnamefont {P.}~\bibnamefont {Lodahl}},\ }\bibfield  {title} {\bibinfo {title} {Deterministic photon{\textendash}emitter coupling in chiral photonic circuits},\ }\href {https://doi.org/10.1038/nnano.2015.159} {\bibfield  {journal} {\bibinfo  {journal}
  {Nature Nanotechnology}\ }\textbf {\bibinfo {volume} {10}},\ \bibinfo {pages} {775} (\bibinfo {year} {2015})}\BibitemShut {NoStop}%
\bibitem [{\citenamefont {Bello}\ \emph {et~al.}(2019)\citenamefont {Bello}, \citenamefont {Platero}, \citenamefont {Cirac},\ and\ \citenamefont {Gonz\'alez-Tudela}}]{doi:10.1126/sciadv.aaw0297}%
  \BibitemOpen
  \bibfield  {author} {\bibinfo {author} {\bibfnamefont {M.}~\bibnamefont {Bello}}, \bibinfo {author} {\bibfnamefont {G.}~\bibnamefont {Platero}}, \bibinfo {author} {\bibfnamefont {J.~I.}\ \bibnamefont {Cirac}},\ and\ \bibinfo {author} {\bibfnamefont {A.}~\bibnamefont {Gonz\'alez-Tudela}},\ }\bibfield  {title} {\bibinfo {title} {Unconventional quantum optics in topological waveguide {QED}},\ }\href {https://doi.org/10.1126/sciadv.aaw0297} {\bibfield  {journal} {\bibinfo  {journal} {Science Advances}\ }\textbf {\bibinfo {volume} {5}},\ \bibinfo {pages} {eaaw0297} (\bibinfo {year} {2019})}\BibitemShut {NoStop}%
\bibitem [{\citenamefont {Liu}\ \emph {et~al.}(2022)\citenamefont {Liu}, \citenamefont {Wang}, \citenamefont {Wang}, \citenamefont {Ma},\ and\ \citenamefont {Cheng}}]{Liu:22}%
  \BibitemOpen
  \bibfield  {author} {\bibinfo {author} {\bibfnamefont {N.}~\bibnamefont {Liu}}, \bibinfo {author} {\bibfnamefont {X.}~\bibnamefont {Wang}}, \bibinfo {author} {\bibfnamefont {X.}~\bibnamefont {Wang}}, \bibinfo {author} {\bibfnamefont {X.-S.}\ \bibnamefont {Ma}},\ and\ \bibinfo {author} {\bibfnamefont {M.-T.}\ \bibnamefont {Cheng}},\ }\bibfield  {title} {\bibinfo {title} {Tunable single photon nonreciprocal scattering based on giant atom-waveguide chiral couplings},\ }\href {https://doi.org/10.1364/OE.460255} {\bibfield  {journal} {\bibinfo  {journal} {Opt. Express}\ }\textbf {\bibinfo {volume} {30}},\ \bibinfo {pages} {23428} (\bibinfo {year} {2022})}\BibitemShut {NoStop}%
\bibitem [{\citenamefont {Mahmoodian}\ \emph {et~al.}(2016)\citenamefont {Mahmoodian}, \citenamefont {Lodahl},\ and\ \citenamefont {S\o{}rensen}}]{PhysRevLett.117.240501}%
  \BibitemOpen
  \bibfield  {author} {\bibinfo {author} {\bibfnamefont {S.}~\bibnamefont {Mahmoodian}}, \bibinfo {author} {\bibfnamefont {P.}~\bibnamefont {Lodahl}},\ and\ \bibinfo {author} {\bibfnamefont {A.~S.}\ \bibnamefont {S\o{}rensen}},\ }\bibfield  {title} {\bibinfo {title} {Quantum networks with chiral-light--matter interaction in waveguides},\ }\href {https://doi.org/10.1103/PhysRevLett.117.240501} {\bibfield  {journal} {\bibinfo  {journal} {Phys. Rev. Lett.}\ }\textbf {\bibinfo {volume} {117}},\ \bibinfo {pages} {240501} (\bibinfo {year} {2016})}\BibitemShut {NoStop}%
\bibitem [{\citenamefont {Mahmoodian}\ \emph {et~al.}(2020)\citenamefont {Mahmoodian}, \citenamefont {Calaj\'o}, \citenamefont {Chang}, \citenamefont {Hammerer},\ and\ \citenamefont {S\o{}rensen}}]{PhysRevX.10.031011}%
  \BibitemOpen
  \bibfield  {author} {\bibinfo {author} {\bibfnamefont {S.}~\bibnamefont {Mahmoodian}}, \bibinfo {author} {\bibfnamefont {G.}~\bibnamefont {Calaj\'o}}, \bibinfo {author} {\bibfnamefont {D.~E.}\ \bibnamefont {Chang}}, \bibinfo {author} {\bibfnamefont {K.}~\bibnamefont {Hammerer}},\ and\ \bibinfo {author} {\bibfnamefont {A.~S.}\ \bibnamefont {S\o{}rensen}},\ }\bibfield  {title} {\bibinfo {title} {Dynamics of many-body photon bound states in chiral waveguide {QED}},\ }\href {https://doi.org/10.1103/PhysRevX.10.031011} {\bibfield  {journal} {\bibinfo  {journal} {Phys. Rev. X}\ }\textbf {\bibinfo {volume} {10}},\ \bibinfo {pages} {031011} (\bibinfo {year} {2020})}\BibitemShut {NoStop}%
\bibitem [{\citenamefont {Leong}\ \emph {et~al.}(2016)\citenamefont {Leong}, \citenamefont {Seidler}, \citenamefont {Steiner}, \citenamefont {Cer\`{e}},\ and\ \citenamefont {Kurtsiefer}}]{Leong2016}%
  \BibitemOpen
  \bibfield  {author} {\bibinfo {author} {\bibfnamefont {V.}~\bibnamefont {Leong}}, \bibinfo {author} {\bibfnamefont {M.~A.}\ \bibnamefont {Seidler}}, \bibinfo {author} {\bibfnamefont {M.}~\bibnamefont {Steiner}}, \bibinfo {author} {\bibfnamefont {A.}~\bibnamefont {Cer\`{e}}},\ and\ \bibinfo {author} {\bibfnamefont {C.}~\bibnamefont {Kurtsiefer}},\ }\bibfield  {title} {\bibinfo {title} {Time-resolved scattering of a single photon by a single atom},\ }\bibfield  {journal} {\bibinfo  {journal} {Nature Communications}\ }\textbf {\bibinfo {volume} {7}},\ \href {https://doi.org/10.1038/ncomms13716} {10.1038/ncomms13716} (\bibinfo {year} {2016})\BibitemShut {NoStop}%
\bibitem [{\citenamefont {Masters}\ \emph {et~al.}(2023)\citenamefont {Masters}, \citenamefont {Hu}, \citenamefont {Cordier}, \citenamefont {Maron}, \citenamefont {Pache}, \citenamefont {Rauschenbeutel}, \citenamefont {Schemmer},\ and\ \citenamefont {Volz}}]{Masters2023}%
  \BibitemOpen
  \bibfield  {author} {\bibinfo {author} {\bibfnamefont {L.}~\bibnamefont {Masters}}, \bibinfo {author} {\bibfnamefont {X.-X.}\ \bibnamefont {Hu}}, \bibinfo {author} {\bibfnamefont {M.}~\bibnamefont {Cordier}}, \bibinfo {author} {\bibfnamefont {G.}~\bibnamefont {Maron}}, \bibinfo {author} {\bibfnamefont {L.}~\bibnamefont {Pache}}, \bibinfo {author} {\bibfnamefont {A.}~\bibnamefont {Rauschenbeutel}}, \bibinfo {author} {\bibfnamefont {M.}~\bibnamefont {Schemmer}},\ and\ \bibinfo {author} {\bibfnamefont {J.}~\bibnamefont {Volz}},\ }\bibfield  {title} {\bibinfo {title} {On the simultaneous scattering of two photons by a single two-level atom},\ }\href {https://doi.org/10.1038/s41566-023-01260-7} {\bibfield  {journal} {\bibinfo  {journal} {Nature Photonics}\ }\textbf {\bibinfo {volume} {17}},\ \bibinfo {pages} {972} (\bibinfo {year} {2023})}\BibitemShut {NoStop}%
\bibitem [{\citenamefont {Tomm}\ \emph {et~al.}(2023)\citenamefont {Tomm}, \citenamefont {Mahmoodian}, \citenamefont {Antoniadis}, \citenamefont {Schott}, \citenamefont {Valentin}, \citenamefont {Wieck}, \citenamefont {Ludwig}, \citenamefont {Javadi},\ and\ \citenamefont {Warburton}}]{Tomm2023}%
  \BibitemOpen
  \bibfield  {author} {\bibinfo {author} {\bibfnamefont {N.}~\bibnamefont {Tomm}}, \bibinfo {author} {\bibfnamefont {S.}~\bibnamefont {Mahmoodian}}, \bibinfo {author} {\bibfnamefont {N.~O.}\ \bibnamefont {Antoniadis}}, \bibinfo {author} {\bibfnamefont {R.}~\bibnamefont {Schott}}, \bibinfo {author} {\bibfnamefont {S.~R.}\ \bibnamefont {Valentin}}, \bibinfo {author} {\bibfnamefont {A.~D.}\ \bibnamefont {Wieck}}, \bibinfo {author} {\bibfnamefont {A.}~\bibnamefont {Ludwig}}, \bibinfo {author} {\bibfnamefont {A.}~\bibnamefont {Javadi}},\ and\ \bibinfo {author} {\bibfnamefont {R.~J.}\ \bibnamefont {Warburton}},\ }\bibfield  {title} {\bibinfo {title} {Photon bound state dynamics from a single artificial atom},\ }\href {https://doi.org/10.1038/s41567-023-01997-6} {\bibfield  {journal} {\bibinfo  {journal} {Nature Physics}\ }\textbf {\bibinfo {volume} {19}},\ \bibinfo {pages} {857} (\bibinfo {year} {2023})}\BibitemShut {NoStop}%
\bibitem [{\citenamefont {Chang}\ \emph {et~al.}(2012)\citenamefont {Chang}, \citenamefont {Jiang}, \citenamefont {Gorshkov},\ and\ \citenamefont {Kimble}}]{Chang_2012}%
  \BibitemOpen
  \bibfield  {author} {\bibinfo {author} {\bibfnamefont {D.~E.}\ \bibnamefont {Chang}}, \bibinfo {author} {\bibfnamefont {L.}~\bibnamefont {Jiang}}, \bibinfo {author} {\bibfnamefont {A.~V.}\ \bibnamefont {Gorshkov}},\ and\ \bibinfo {author} {\bibfnamefont {H.~J.}\ \bibnamefont {Kimble}},\ }\bibfield  {title} {\bibinfo {title} {Cavity {QED} with atomic mirrors},\ }\href {https://doi.org/10.1088/1367-2630/14/6/063003} {\bibfield  {journal} {\bibinfo  {journal} {New Journal of Physics}\ }\textbf {\bibinfo {volume} {14}},\ \bibinfo {pages} {063003} (\bibinfo {year} {2012})}\BibitemShut {NoStop}%
\bibitem [{\citenamefont {Droenner}\ \emph {et~al.}(2019)\citenamefont {Droenner}, \citenamefont {Naumann}, \citenamefont {Sch\"oll}, \citenamefont {Knorr},\ and\ \citenamefont {Carmele}}]{Droenner2019}%
  \BibitemOpen
  \bibfield  {author} {\bibinfo {author} {\bibfnamefont {L.}~\bibnamefont {Droenner}}, \bibinfo {author} {\bibfnamefont {N.~L.}\ \bibnamefont {Naumann}}, \bibinfo {author} {\bibfnamefont {E.}~\bibnamefont {Sch\"oll}}, \bibinfo {author} {\bibfnamefont {A.}~\bibnamefont {Knorr}},\ and\ \bibinfo {author} {\bibfnamefont {A.}~\bibnamefont {Carmele}},\ }\bibfield  {title} {\bibinfo {title} {Quantum {Pyragas} control: {Selective} control of individual photon probabilities},\ }\href {https://doi.org/10.1103/PhysRevA.99.023840} {\bibfield  {journal} {\bibinfo  {journal} {Physical Review A}\ }\textbf {\bibinfo {volume} {99}},\ \bibinfo {pages} {023840} (\bibinfo {year} {2019})}\BibitemShut {NoStop}%
\bibitem [{\citenamefont {Arranz~Regidor}\ \emph {et~al.}(2021)\citenamefont {Arranz~Regidor}, \citenamefont {Crowder}, \citenamefont {Carmichael},\ and\ \citenamefont {Hughes}}]{PhysRevResearch.3.023030}%
  \BibitemOpen
  \bibfield  {author} {\bibinfo {author} {\bibfnamefont {S.}~\bibnamefont {Arranz~Regidor}}, \bibinfo {author} {\bibfnamefont {G.}~\bibnamefont {Crowder}}, \bibinfo {author} {\bibfnamefont {H.}~\bibnamefont {Carmichael}},\ and\ \bibinfo {author} {\bibfnamefont {S.}~\bibnamefont {Hughes}},\ }\bibfield  {title} {\bibinfo {title} {Modeling quantum light-matter interactions in waveguide {QED} with retardation, nonlinear interactions, and a time-delayed feedback: Matrix product states versus a space-discretized waveguide model},\ }\href {https://doi.org/10.1103/PhysRevResearch.3.023030} {\bibfield  {journal} {\bibinfo  {journal} {Phys. Rev. Res.}\ }\textbf {\bibinfo {volume} {3}},\ \bibinfo {pages} {023030} (\bibinfo {year} {2021})}\BibitemShut {NoStop}%
\bibitem [{\citenamefont {Arranz~Regidor}\ \emph {et~al.}(2025)\citenamefont {Arranz~Regidor}, \citenamefont {Knorr},\ and\ \citenamefont {Hughes}}]{sofia2025}%
  \BibitemOpen
  \bibfield  {author} {\bibinfo {author} {\bibfnamefont {S.}~\bibnamefont {Arranz~Regidor}}, \bibinfo {author} {\bibfnamefont {A.}~\bibnamefont {Knorr}},\ and\ \bibinfo {author} {\bibfnamefont {S.}~\bibnamefont {Hughes}},\ }\bibfield  {title} {\bibinfo {title} {Theory and simulations of few-photon {Fock} state pulses strongly interacting with a single qubit in a waveguide: {Exact} population dynamics and time-dependent spectra},\ }\href {https://doi.org/10.1103/lp1b-yswm} {\bibfield  {journal} {\bibinfo  {journal} {Physical Review Research}\ }\textbf {\bibinfo {volume} {7}},\ \bibinfo {pages} {23295} (\bibinfo {year} {2025})}\BibitemShut {NoStop}%
\bibitem [{\citenamefont {Fan}\ \emph {et~al.}(2010)\citenamefont {Fan}, \citenamefont {Kocaba\mbox{\c{s}}},\ and\ \citenamefont {Shen}}]{PhysRevA.82.063821}%
  \BibitemOpen
  \bibfield  {author} {\bibinfo {author} {\bibfnamefont {S.}~\bibnamefont {Fan}}, \bibinfo {author} {\bibfnamefont {S.~E.}\ \bibnamefont {Kocaba\mbox{\c{s}}}},\ and\ \bibinfo {author} {\bibfnamefont {J.-T.}\ \bibnamefont {Shen}},\ }\bibfield  {title} {\bibinfo {title} {Input-output formalism for few-photon transport in one-dimensional nanophotonic waveguides coupled to a qubit},\ }\href {https://doi.org/10.1103/PhysRevA.82.063821} {\bibfield  {journal} {\bibinfo  {journal} {Phys. Rev. A}\ }\textbf {\bibinfo {volume} {82}},\ \bibinfo {pages} {063821} (\bibinfo {year} {2010})}\BibitemShut {NoStop}%
\bibitem [{\citenamefont {Rephaeli}\ and\ \citenamefont {Fan}(2012)}]{Rephaeli2012FewPhotonSC}%
  \BibitemOpen
  \bibfield  {author} {\bibinfo {author} {\bibfnamefont {E.}~\bibnamefont {Rephaeli}}\ and\ \bibinfo {author} {\bibfnamefont {S.}~\bibnamefont {Fan}},\ }\bibfield  {title} {\bibinfo {title} {Few-photon single-atom cavity {QED} with input-output formalism in fock space},\ }\href {https://api.semanticscholar.org/CorpusID:40913017} {\bibfield  {journal} {\bibinfo  {journal} {IEEE Journal of Selected Topics in Quantum Electronics}\ }\textbf {\bibinfo {volume} {18}},\ \bibinfo {pages} {1754} (\bibinfo {year} {2012})}\BibitemShut {NoStop}%
\bibitem [{\citenamefont {Wang}\ \emph {et~al.}(2011)\citenamefont {Wang}, \citenamefont {Min\'a\ifmmode~\check{r}\else \v{r}\fi{}}, \citenamefont {Sheridan},\ and\ \citenamefont {Scarani}}]{PhysRevA.83.063842}%
  \BibitemOpen
  \bibfield  {author} {\bibinfo {author} {\bibfnamefont {Y.}~\bibnamefont {Wang}}, \bibinfo {author} {\bibfnamefont {J.~c.~v.}\ \bibnamefont {Min\'a\ifmmode~\check{r}\else \v{r}\fi{}}}, \bibinfo {author} {\bibfnamefont {L.}~\bibnamefont {Sheridan}},\ and\ \bibinfo {author} {\bibfnamefont {V.}~\bibnamefont {Scarani}},\ }\bibfield  {title} {\bibinfo {title} {Efficient excitation of a two-level atom by a single photon in a propagating mode},\ }\href {https://doi.org/10.1103/PhysRevA.83.063842} {\bibfield  {journal} {\bibinfo  {journal} {Phys. Rev. A}\ }\textbf {\bibinfo {volume} {83}},\ \bibinfo {pages} {063842} (\bibinfo {year} {2011})}\BibitemShut {NoStop}%
\bibitem [{\citenamefont {Barkemeyer}\ \emph {et~al.}(2022)\citenamefont {Barkemeyer}, \citenamefont {Knorr},\ and\ \citenamefont {Carmele}}]{PhysRevA.106.023708}%
  \BibitemOpen
  \bibfield  {author} {\bibinfo {author} {\bibfnamefont {K.}~\bibnamefont {Barkemeyer}}, \bibinfo {author} {\bibfnamefont {A.}~\bibnamefont {Knorr}},\ and\ \bibinfo {author} {\bibfnamefont {A.}~\bibnamefont {Carmele}},\ }\bibfield  {title} {\bibinfo {title} {Heisenberg treatment of multiphoton pulses in waveguide {QED} with time-delayed feedback},\ }\href {https://doi.org/10.1103/PhysRevA.106.023708} {\bibfield  {journal} {\bibinfo  {journal} {Phys. Rev. A}\ }\textbf {\bibinfo {volume} {106}},\ \bibinfo {pages} {023708} (\bibinfo {year} {2022})}\BibitemShut {NoStop}%
\bibitem [{\citenamefont {Nysteen}\ \emph {et~al.}(2015)\citenamefont {Nysteen}, \citenamefont {Kristensen}, \citenamefont {McCutcheon}, \citenamefont {Kaer},\ and\ \citenamefont {M{\o}rk}}]{Nysteen2015}%
  \BibitemOpen
  \bibfield  {author} {\bibinfo {author} {\bibfnamefont {A.}~\bibnamefont {Nysteen}}, \bibinfo {author} {\bibfnamefont {P.~T.}\ \bibnamefont {Kristensen}}, \bibinfo {author} {\bibfnamefont {D.~P.~S.}\ \bibnamefont {McCutcheon}}, \bibinfo {author} {\bibfnamefont {P.}~\bibnamefont {Kaer}},\ and\ \bibinfo {author} {\bibfnamefont {J.}~\bibnamefont {M{\o}rk}},\ }\bibfield  {title} {\bibinfo {title} {Scattering of two photons on a quantum emitter in a one-dimensional waveguide: exact dynamics and induced correlations},\ }\href {https://doi.org/10.1088/1367-2630/17/2/023030} {\bibfield  {journal} {\bibinfo  {journal} {New Journal of Physics}\ }\textbf {\bibinfo {volume} {17}},\ \bibinfo {pages} {023030} (\bibinfo {year} {2015})}\BibitemShut {NoStop}%
\bibitem [{\citenamefont {Chen}\ \emph {et~al.}(2011)\citenamefont {Chen}, \citenamefont {Wubs}, \citenamefont {M{\o}rk},\ and\ \citenamefont {Koenderink}}]{Chen_2011}%
  \BibitemOpen
  \bibfield  {author} {\bibinfo {author} {\bibfnamefont {Y.}~\bibnamefont {Chen}}, \bibinfo {author} {\bibfnamefont {M.}~\bibnamefont {Wubs}}, \bibinfo {author} {\bibfnamefont {J.}~\bibnamefont {M{\o}rk}},\ and\ \bibinfo {author} {\bibfnamefont {A.~F.}\ \bibnamefont {Koenderink}},\ }\bibfield  {title} {\bibinfo {title} {Coherent single-photon absorption by single emitters coupled to one-dimensional nanophotonic waveguides},\ }\href {https://doi.org/10.1088/1367-2630/13/10/103010} {\bibfield  {journal} {\bibinfo  {journal} {New Journal of Physics}\ }\textbf {\bibinfo {volume} {13}},\ \bibinfo {pages} {103010} (\bibinfo {year} {2011})}\BibitemShut {NoStop}%
\bibitem [{\citenamefont {Domokos}\ \emph {et~al.}(2002)\citenamefont {Domokos}, \citenamefont {Horak},\ and\ \citenamefont {Ritsch}}]{PhysRevA.65.033832}%
  \BibitemOpen
  \bibfield  {author} {\bibinfo {author} {\bibfnamefont {P.}~\bibnamefont {Domokos}}, \bibinfo {author} {\bibfnamefont {P.}~\bibnamefont {Horak}},\ and\ \bibinfo {author} {\bibfnamefont {H.}~\bibnamefont {Ritsch}},\ }\bibfield  {title} {\bibinfo {title} {Quantum description of light-pulse scattering on a single atom in waveguides},\ }\href {https://doi.org/10.1103/PhysRevA.65.033832} {\bibfield  {journal} {\bibinfo  {journal} {Phys. Rev. A}\ }\textbf {\bibinfo {volume} {65}},\ \bibinfo {pages} {033832} (\bibinfo {year} {2002})}\BibitemShut {NoStop}%
\bibitem [{\citenamefont {Roulet}\ \emph {et~al.}(2016)\citenamefont {Roulet}, \citenamefont {Le},\ and\ \citenamefont {Scarani}}]{PhysRevA.93.033838}%
  \BibitemOpen
  \bibfield  {author} {\bibinfo {author} {\bibfnamefont {A.}~\bibnamefont {Roulet}}, \bibinfo {author} {\bibfnamefont {H.~N.}\ \bibnamefont {Le}},\ and\ \bibinfo {author} {\bibfnamefont {V.}~\bibnamefont {Scarani}},\ }\bibfield  {title} {\bibinfo {title} {Two photons on an atomic beam splitter: Nonlinear scattering and induced correlations},\ }\href {https://doi.org/10.1103/PhysRevA.93.033838} {\bibfield  {journal} {\bibinfo  {journal} {Phys. Rev. A}\ }\textbf {\bibinfo {volume} {93}},\ \bibinfo {pages} {033838} (\bibinfo {year} {2016})}\BibitemShut {NoStop}%
\bibitem [{\citenamefont {Stolyarov}(2019)}]{PhysRevA.99.023857}%
  \BibitemOpen
  \bibfield  {author} {\bibinfo {author} {\bibfnamefont {E.~V.}\ \bibnamefont {Stolyarov}},\ }\bibfield  {title} {\bibinfo {title} {Few-photon fock-state wave packet interacting with a cavity-atom system in a waveguide: Exact quantum state dynamics},\ }\href {https://doi.org/10.1103/PhysRevA.99.023857} {\bibfield  {journal} {\bibinfo  {journal} {Phys. Rev. A}\ }\textbf {\bibinfo {volume} {99}},\ \bibinfo {pages} {023857} (\bibinfo {year} {2019})}\BibitemShut {NoStop}%
\bibitem [{\citenamefont {Chumak}\ and\ \citenamefont {Stolyarov}(2014)}]{PhysRevA.90.063832}%
  \BibitemOpen
  \bibfield  {author} {\bibinfo {author} {\bibfnamefont {O.~O.}\ \bibnamefont {Chumak}}\ and\ \bibinfo {author} {\bibfnamefont {E.~V.}\ \bibnamefont {Stolyarov}},\ }\bibfield  {title} {\bibinfo {title} {Photon distribution function for propagation of two-photon pulses in waveguide-qubit systems},\ }\href {https://doi.org/10.1103/PhysRevA.90.063832} {\bibfield  {journal} {\bibinfo  {journal} {Phys. Rev. A}\ }\textbf {\bibinfo {volume} {90}},\ \bibinfo {pages} {063832} (\bibinfo {year} {2014})}\BibitemShut {NoStop}%
\bibitem [{\citenamefont {McCulloch}(2007)}]{mcculloch_density-matrix_2007}%
  \BibitemOpen
  \bibfield  {author} {\bibinfo {author} {\bibfnamefont {I.~P.}\ \bibnamefont {McCulloch}},\ }\bibfield  {title} {\bibinfo {title} {From density-matrix renormalization group to matrix product states},\ }\href {https://doi.org/10.1088/1742-5468/2007/10/p10014} {\bibfield  {journal} {\bibinfo  {journal} {Journal of Statistical Mechanics: Theory and Experiment}\ }\textbf {\bibinfo {volume} {2007}},\ \bibinfo {pages} {P10014} (\bibinfo {year} {2007})}\BibitemShut {NoStop}%
\bibitem [{\citenamefont {Or\'us}(2014)}]{orus_practical_2014}%
  \BibitemOpen
  \bibfield  {author} {\bibinfo {author} {\bibfnamefont {R.}~\bibnamefont {Or\'us}},\ }\bibfield  {title} {\bibinfo {title} {A practical introduction to tensor networks: Matrix product states and projected entangled pair states},\ }\href {https://doi.org/10.1016/j.aop.2014.06.013} {\bibfield  {journal} {\bibinfo  {journal} {Annals of Physics}\ }\textbf {\bibinfo {volume} {349}},\ \bibinfo {pages} {117} (\bibinfo {year} {2014})}\BibitemShut {NoStop}%
\bibitem [{\citenamefont {Barkemeyer}\ \emph {et~al.}(2021)\citenamefont {Barkemeyer}, \citenamefont {Knorr},\ and\ \citenamefont {Carmele}}]{PhysRevA.103.033704}%
  \BibitemOpen
  \bibfield  {author} {\bibinfo {author} {\bibfnamefont {K.}~\bibnamefont {Barkemeyer}}, \bibinfo {author} {\bibfnamefont {A.}~\bibnamefont {Knorr}},\ and\ \bibinfo {author} {\bibfnamefont {A.}~\bibnamefont {Carmele}},\ }\bibfield  {title} {\bibinfo {title} {Strongly entangled system-reservoir dynamics with multiphoton pulses beyond the two-excitation limit: Exciting the atom-photon bound state},\ }\href {https://doi.org/10.1103/PhysRevA.103.033704} {\bibfield  {journal} {\bibinfo  {journal} {Phys. Rev. A}\ }\textbf {\bibinfo {volume} {103}},\ \bibinfo {pages} {033704} (\bibinfo {year} {2021})}\BibitemShut {NoStop}%
\bibitem [{\citenamefont {Regidor}\ \emph {et~al.}(2026)\citenamefont {Regidor}, \citenamefont {Kozma},\ and\ \citenamefont {Hughes}}]{2602.15826}%
  \BibitemOpen
  \bibfield  {author} {\bibinfo {author} {\bibfnamefont {S.~A.}\ \bibnamefont {Regidor}}, \bibinfo {author} {\bibfnamefont {M.}~\bibnamefont {Kozma}},\ and\ \bibinfo {author} {\bibfnamefont {S.}~\bibnamefont {Hughes}},\ }\href@noop {} {\bibinfo {title} {{QwaveMPS}: An efficient open-source python package for simulating {non-Markovian waveguide-QED} using matrix product states}} (\bibinfo {year} {2026}),\ \Eprint {https://arxiv.org/abs/arXiv:2602.15826} {arXiv:2602.15826} \BibitemShut {NoStop}%
\bibitem [{\citenamefont {Ramos}\ and\ \citenamefont {Garc\'{\i}a-Ripoll}(2017)}]{PhysRevLett.119.153601}%
  \BibitemOpen
  \bibfield  {author} {\bibinfo {author} {\bibfnamefont {T.}~\bibnamefont {Ramos}}\ and\ \bibinfo {author} {\bibfnamefont {J.~J.}\ \bibnamefont {Garc\'{\i}a-Ripoll}},\ }\bibfield  {title} {\bibinfo {title} {Multiphoton scattering tomography with coherent states},\ }\href {https://doi.org/10.1103/PhysRevLett.119.153601} {\bibfield  {journal} {\bibinfo  {journal} {Phys. Rev. Lett.}\ }\textbf {\bibinfo {volume} {119}},\ \bibinfo {pages} {153601} (\bibinfo {year} {2017})}\BibitemShut {NoStop}%
\bibitem [{\citenamefont {Gardiner}\ and\ \citenamefont {Zoller}(2010)}]{gardiner_zoller_2010}%
  \BibitemOpen
  \bibfield  {author} {\bibinfo {author} {\bibfnamefont {C.~W.}\ \bibnamefont {Gardiner}}\ and\ \bibinfo {author} {\bibfnamefont {P.}~\bibnamefont {Zoller}},\ }\href@noop {} {\emph {\bibinfo {title} {Quantum noise: a handbook of {Markovian} and {non-Markovian} quantum stochastic methods with applications to quantum optics}}}\ (\bibinfo  {publisher} {Springer},\ \bibinfo {address} {Berlin},\ \bibinfo {year} {2010})\BibitemShut {NoStop}%
\bibitem [{\citenamefont {Guimond}\ \emph {et~al.}(2017)\citenamefont {Guimond}, \citenamefont {Pletyukhov}, \citenamefont {Pichler},\ and\ \citenamefont {Zoller}}]{Guimond_2017}%
  \BibitemOpen
  \bibfield  {author} {\bibinfo {author} {\bibfnamefont {P.-O.}\ \bibnamefont {Guimond}}, \bibinfo {author} {\bibfnamefont {M.}~\bibnamefont {Pletyukhov}}, \bibinfo {author} {\bibfnamefont {H.}~\bibnamefont {Pichler}},\ and\ \bibinfo {author} {\bibfnamefont {P.}~\bibnamefont {Zoller}},\ }\bibfield  {title} {\bibinfo {title} {Delayed coherent quantum feedback from a scattering theory and a matrix product state perspective},\ }\href {https://doi.org/10.1088/2058-9565/aa7f03} {\bibfield  {journal} {\bibinfo  {journal} {Quantum Science and Technology}\ }\textbf {\bibinfo {volume} {2}},\ \bibinfo {pages} {044012} (\bibinfo {year} {2017})}\BibitemShut {NoStop}%
\bibitem [{\citenamefont {Mukhopadhyay}\ and\ \citenamefont {Shen}(2024)}]{Mukhopadhyay_2024}%
  \BibitemOpen
  \bibfield  {author} {\bibinfo {author} {\bibfnamefont {D.}~\bibnamefont {Mukhopadhyay}}\ and\ \bibinfo {author} {\bibfnamefont {J.-T.}\ \bibnamefont {Shen}},\ }\bibfield  {title} {\bibinfo {title} {Quantum multiphoton {R}abi oscillations in waveguide {QED}},\ }\href {https://doi.org/10.1088/1367-2630/ad80b7} {\bibfield  {journal} {\bibinfo  {journal} {New Journal of Physics}\ }\textbf {\bibinfo {volume} {26}},\ \bibinfo {pages} {103026} (\bibinfo {year} {2024})}\BibitemShut {NoStop}%
\bibitem [{\citenamefont {Chumak}\ and\ \citenamefont {Stolyarov}(2013)}]{PhysRevA.88.013855}%
  \BibitemOpen
  \bibfield  {author} {\bibinfo {author} {\bibfnamefont {O.~O.}\ \bibnamefont {Chumak}}\ and\ \bibinfo {author} {\bibfnamefont {E.~V.}\ \bibnamefont {Stolyarov}},\ }\bibfield  {title} {\bibinfo {title} {Phase-space distribution functions for photon propagation in waveguides coupled to a qubit},\ }\href {https://doi.org/10.1103/PhysRevA.88.013855} {\bibfield  {journal} {\bibinfo  {journal} {Phys. Rev. A}\ }\textbf {\bibinfo {volume} {88}},\ \bibinfo {pages} {013855} (\bibinfo {year} {2013})}\BibitemShut {NoStop}%
\bibitem [{\citenamefont {Bundgaard-Nielsen}\ \emph {et~al.}(2026)\citenamefont {Bundgaard-Nielsen}, \citenamefont {Kozma}, \citenamefont {Regidor},\ and\ \citenamefont {Hughes}}]{bundgaardnielsen2026}%
  \BibitemOpen
  \bibfield  {author} {\bibinfo {author} {\bibfnamefont {M.}~\bibnamefont {Bundgaard-Nielsen}}, \bibinfo {author} {\bibfnamefont {M.}~\bibnamefont {Kozma}}, \bibinfo {author} {\bibfnamefont {S.~A.}\ \bibnamefont {Regidor}},\ and\ \bibinfo {author} {\bibfnamefont {S.}~\bibnamefont {Hughes}},\ }\href {https://arxiv.org/abs/2604.18036} {\bibinfo {title} {Decoherence in waveguide quantum electrodynamics using matrix product states}} (\bibinfo {year} {2026}),\ \Eprint {https://arxiv.org/abs/2604.18036} {arXiv:2604.18036 [quant-ph]} \BibitemShut {NoStop}%
\bibitem [{\citenamefont {Crosswhite}\ and\ \citenamefont {Bacon}(2008)}]{bacon2008nfa}%
  \BibitemOpen
  \bibfield  {author} {\bibinfo {author} {\bibfnamefont {G.~M.}\ \bibnamefont {Crosswhite}}\ and\ \bibinfo {author} {\bibfnamefont {D.}~\bibnamefont {Bacon}},\ }\bibfield  {title} {\bibinfo {title} {Finite automata for caching in matrix product algorithms},\ }\href {https://doi.org/10.1103/PhysRevA.78.012356} {\bibfield  {journal} {\bibinfo  {journal} {Phys. Rev. A}\ }\textbf {\bibinfo {volume} {78}},\ \bibinfo {pages} {012356} (\bibinfo {year} {2008})}\BibitemShut {NoStop}%
\end{thebibliography}%

\end{document}